\documentclass[12pt]{article}
\usepackage{latexsym}
\usepackage{ifthen}
\usepackage{epsf}
\def\hybrid{\topmargin -20pt    \oddsidemargin 0pt
        \headheight 0pt \headsep 0pt
        \textwidth 6.5in        
        \textheight 9in         
        \marginparwidth .875in
        \parskip 5pt plus 1pt   \jot = 1.5ex}

\hybrid
\makeatletter
\@addtoreset{equation}{section}
\makeatother

\newcommand{\color}[6]{}

\newcommand{\cD}{{\cal D}}

\newcommand{\cL}{{\cal L}}

\newcommand{\cN}{{\cal N}}
\newcommand{\cO}{{\cal O}}

\newcommand{\hf}{\frac12}
\newcommand{\qt}{\frac14}
\newcommand{\bea}{\begin{eqnarray}}
\newcommand{\eea}{\end{eqnarray}}
\newcommand{\be}{\begin{equation}}
\newcommand{\ee}{\end{equation}}
\newcommand{\bt}{\begin{tabular}}
\newcommand{\et}{\end{tabular}}
\newcommand{\ba}{\begin{array}}
\newcommand{\ea}{\end{array}}

\newcommand{\tr}{\mathop{\rm tr}}

\newcommand{\diag}{{\rm diag}}
\newcommand{\R}{{\rm Re}}
\newcommand{\I}{{\rm Im}}

\newcommand{\Mpl}{M_{\rm Pl}}
\def\IB{\relax{\rm I\kern-.18em B}}
\def\ID{\relax{\rm I\kern-.18em D}}
\def\IE{\relax{\rm I\kern-.18em E}}
\def\IF{\relax{\rm I\kern-.18em F}}
\def\IH{\relax{\rm I\kern-.18em H}}
\def\II{\relax{\rm I\kern-.18em I}}
\def\IK{\relax{\rm I\kern-.18em K}}
\def\IL{\relax{\rm I\kern-.18em L}}
\def\IM{\relax{\rm I\kern-.18em M}}
\def\IN{\relax{\rm I\kern-.18em N}}
\def\IP{\relax{\rm I\kern-.18em P}}
\def\IR{\relax{\rm I\kern-.18em R}}
\def\IT{\relax{\rm I\kern-.42em T}}
\def\IZ{\relax{\hbox{\raisebox{.38ex}
    {\scriptsize\bfseries\slshape /}\kern-.40em\_\kern-.28em\rm Z}}}
\def\Iz{\relax{\hbox{\raisebox{.38ex}
    {\tiny\bfseries\slshape /}\kern-.25em\raisebox{.65ex}
    {\tiny\bfseries\slshape /}\kern-.43em\_\kern-.26em\rm Z}}}
\def\inbar{\vrule height1.5ex width.8pt depth-0.2pt}
\def\inbarhi{\vrule height1.55ex width.5pt depth-.85ex}
\def\inbarlo{\vrule height.8ex width.5pt depth0ex}
\def\IC{\relax{\rm C\kern-.48em \inbar\kern.48em}}
\def\IO{\relax{\rm O\kern-.56em \inbar\kern.56em}}
\def\IQ{\relax{\rm Q\kern-.56em \inbar\kern.56em}}
\def\IS{\relax{\rm S\kern-.37em \inbarhi\kern.08em\inbarlo\kern.29em}}

\def \one{\relax{\rm 1\kern-.26em I}}

\def \soll={\stackrel{!}{=}}

\def \Mpl{M_{\rm Pl}}
\def \Mstr{M_{\rm str}}
\def \Mc{M_{\rm c}}

\def \ms{m_{\rm S}}

\def \dm{\partial_m}

\def \Dm{\cD_m}

\def \DDm{\cD^m}

\def \DM{\cD_M}

\def \DDM{\cD^M}

\def \eps{\epsilon}

\def \first{$1^{\rm st}$}
\def \second{$2^{\rm nd}$}

\def \barfill{\leaders\hrule height 0.1 true pt\hfill}
\def \overbar#1{\vbox{\ialign{##\crcr\barfill\crcr\noalign{\kern 1pt
                                      \nointerlineskip}$\hfil{#1}\hfil$\crcr}}}
\def \scriptbar#1{{\vbox{\ialign{##\crcr\thinspace\barfill\thinspace\crcr
    \noalign{\kern 0.8pt\nointerlineskip}$\hfil{\scriptstyle #1}\hfil$\crcr}}}}

\def \wbar#1{\overline{#1}}

\newlength{\oldindent}
\newlength{\quadlength} 
\newlength{\abstand} \newlength{\breite}

\def\cnodea{\put(0.8,2.5){\circle*{1.6}}}
\def\cnodeb{\put(7.4,2.5){\circle*{1.6}}}
\def\rnodea{\put(0.8,2.5){\circle{1.6}}}
\def\rnodeb{\put(7.4,2.5){\circle{1.6}}}
\def\pnodea{\put(0.8,2.5){\circle{1.6}}\put(0.8,2.5){\circle*{0.4}}}
\def\pnodeb{\put(7.4,2.5){\circle{1.6}}\put(7.4,2.5){\circle*{0.4}}}
\newlength{\firstlength} \newlength{\secondlength}
\def\qlink#1#2{\begin{picture}(9,4)
            \def\first{#1} \def\second{#2}
            \settowidth{\firstlength}{$k$} \settowidth{\secondlength}{$l$}
            \addtolength{\firstlength}{-0.5\firstlength}
            \addtolength{\secondlength}{-0.5\secondlength}
            \def\cx{c} \def\rx{r} \def\px{p}
            \ifx\first\cx \cnodea \else
               \ifx\first\rx \rnodea \else \pnodea \fi\fi
            \ifx\second\cx \put(1.6,2.5){\vector(1,0){5}}\cnodeb \else
               \put(1.6,2.5){\line(1,0){5}} \ifx\second\rx \rnodeb \else
               \pnodeb\fi\fi
            \put(1,-0.9){\hspace{-\firstlength}\scriptsize$k$}
            \put(7.6,-0.9){\hspace{-\secondlength}\scriptsize$l$}
            \end{picture}}
\def\qlinkx#1#2#3#4#5#6{\begin{picture}(9,4)
            \def\tail{#1} \def\head{#2}
            \def\first{#3} \def\second{#4}
            \settowidth{\firstlength}{$#5$} \settowidth{\secondlength}{$#6$}
            \addtolength{\firstlength}{-0.5\firstlength}
            \addtolength{\secondlength}{-0.5\secondlength}
            \def\cx{c} \def\rx{r} \def\px{p} \def\yes{1} \def\no{0}
            \ifx\first\cx \cnodea \else
               \ifx\first\rx \rnodea \else \pnodea \fi\fi
            \if\head\yes \put(1.6,2.5){\vector(1,0){5}} 
               \else\ifx\tail\no \put(1.6,2.5){\line(1,0){5}} \fi\fi
            \if\tail\yes \put(6.6,2.5){\vector(-1,0){5}} \fi
            \ifx\second\cx \cnodeb \else
               \ifx\second\rx \rnodeb \else \pnodeb\fi\fi
            \put(1,-0.9){\hspace{-\firstlength}\scriptsize$#5$}
            \put(7.6,-0.9){\hspace{-\secondlength}\scriptsize$#6$}
            \end{picture}}
\def\qtens#1#2{\def\first{#1}\def\cx{c}\def\rx{r}\def\px{p}%
               \settowidth{\firstlength}{$#2$}
               \addtolength{\firstlength}{-0.5\firstlength}
             \ifx\first\cx 
                   \begin{picture}(3,6)
                     \cnodea
                     \put(0.8,4.5){\circle{4}}
                     \put(1,-0.9){\hspace{-\firstlength}\scriptsize$#2$}
                   \end{picture}%
              \else%
                   \begin{picture}(3,6)
                     \ifx\first\rx\rnodea\else\pnodea\fi
                     \qbezier(-0.1,3)(-1.2,3.72843)(-1.2,4.5)
                     \qbezier(-1.2,4.5)(-1.2,5.32843)(-0.61421,5.91421)
                     \qbezier(-0.61421,5.91421)(-0.02843,6.5)(0.8,6.5)
                     \qbezier(0.8,6.5)(1.62843,6.5)(2.21421,5.91421)
                     \qbezier(2.21421,5.91421)(2.8,5.32843)(2.8,4.5)
                     \qbezier(2.8,4.5)(2.8,3.6)(1.6,3)
                     \put(1,-0.9){\hspace{-\firstlength}\scriptsize$#2$}
                   \end{picture}%
              \fi}

\def\fpropagator#1#2{\begin{picture}(20,5)
            \def\first{#1} \def\second{#2}
            \settowidth{\firstlength}{$#1$} \settowidth{\secondlength}{$#2$}
            \put(0,-2){$#1$}
            \put(20,2){\vector(-1,0){10}}\put(10,2){\line(-1,0){10}}
            \put(20,-2){\hspace{-\secondlength}$#2$}
            \end{picture}}

\def\fvertex#1#2#3{\begin{picture}(12,10)
            \def\first{#1} \def\second{#2}
            \settowidth{\firstlength}{$#1$} \settowidth{\secondlength}{$#2$}
            \put(6,5){\circle*{1}}
            \put(0,-2){$#1$}
            \put(6,5){\line(-5,-3){6}}\put(6,5){\line(5,-3){6}}
            \put(6,5.5){\line(0,1){1.5}}\put(6,8){\line(0,1){1.5}}
            \put(6,10.5){\line(0,1){1.5}}
            \put(12,-2){\hspace{-\secondlength}$#2$}
            \put(7,9){$#3$}
            \end{picture}}

\newlength{\totallength}
\newlength{\height}
\newlength{\firstheight}\newlength{\secondheight}
\newlength{\ruleheight}\newlength{\midruleheight}
\newlength{\lineheight}\newlength{\hcorrection}
\newcommand{\contraction}[3][0em]{\setlength{\hcorrection}{#1}
                \settowidth{\firstlength}{$#2$} 
                \settowidth{\secondlength}{$#3$}
                \settoheight{\firstheight}{$#2$}
                \settoheight{\secondheight}{$#3$}
                \setlength{\lineheight}{0.3pt}
                \setlength{\ruleheight}{1ex}
                \setlength{\midruleheight}{\ruleheight}
                \addtolength{\midruleheight}{-\lineheight}
                \addtolength{\firstlength}{-0.5\firstlength}
                \addtolength{\secondlength}{-0.5\secondlength}
                \setlength{\totallength}{\firstlength}
                \addtolength{\totallength}{\secondlength}
                \ifthenelse{\firstheight<\secondheight}
                           {\setlength{\height}{\secondheight}}
                           {\setlength{\height}{\firstheight}}
                \addtolength{\height}{0.25ex}
                \hspace*{\firstlength}\hspace*{\hcorrection}
                \rule[\height]{\lineheight}{\ruleheight}
                \addtolength{\height}{\midruleheight}
                \rule[\height]{\totallength}{\lineheight}
                \addtolength{\height}{-\midruleheight}
                \rule[\height]{\lineheight}{\ruleheight}
                \hspace{-\firstlength}\hspace{-\totallength}
                \hspace*{-\hcorrection}
                #2#3}

\def \Nucl#1{{\em Nucl.~Phys.}\ {\bf B#1}}

\def \PhysR#1{{\em Phys.~Rev.}\ {\bf D#1}}

\def \PhysRev#1{{\em Phys.~Rev.}\ {\bf #1}}
\def \PhysRep#1{{\em Phys.~Rep.}\ {\bf #1}}
\def \PhysL#1{{\em Phys.~Lett.}\ {\bf #1B}}

\def \IntJ#1{{\em Int.~J.~Mod.~Phys.}\ {\bf A#1}}

\def \Vm{V^m_{\xi_2}}

\def\oneloop{\hbox{\scriptsize 1-loop}}
\def\nabl{\nabla_{\!\!5}}

\renewcommand{\thefootnote}{\fnsymbol{footnote}}

\begin{document}

\begin{titlepage}
\begin{center}

\rightline{}
\rightline{SLAC-PUB-9521}
\rightline{hep-th/0209206}

\vskip .6in
{\LARGE \bf Loop-Effects in Pseudo-Supersymmetry}
\vskip .8in

{\bf Matthias Klein\footnote{E-mail: mklein@slac.stanford.edu}}
\vskip 0.8cm
{\em SLAC, Stanford University, Stanford, CA 94309, USA.}

\end{center}

\vskip 1.5cm

\begin{center} {\bf ABSTRACT } \end{center}
We analyze the transmission of supersymmetry breaking in brane-world
models of pseudo-supersymmetry. In these models two branes preserve
different halves of the bulk supersymmetry. Thus supersymmetry is 
broken although each sector of the model is supersymmetric when 
considered separately. The world-volume theory on one brane feels
the breakdown of supersymmetry only through two-loop interactions
involving a coupling to fields from the other brane. 
In a 5D toy model with bulk vectors,
we compute the diagrams that contribute to scalar masses on one
brane and find that the masses are proportional to the compactification 
scale up to logarithmic corrections, 
$m^2\propto (2\pi R)^{-2}(\ln(2\pi R\,\ms)-1.1)$, where $\ms$ is an
ultraviolet cutoff.
Thus, for large compactification radii, where this result is valid,
the brane scalars acquire a positive mass squared. 
We also compute the three-loop diagrams relevant to the Casimir energy
between the two branes and find
$E\propto (2\pi R)^{-4}\Big((\ln(2\pi R\,\ms)-1.7)^2+0.2\Big)$.
For large radii, this yields a repulsive Casimir force.

\vfill

September 2002
\end{titlepage}

\newpage

\setcounter{page}{1} \pagestyle{plain}
\renewcommand{\thefootnote}{\arabic{footnote}}
\setcounter{footnote}{0}


\section{Introduction}
Supersymmetry breaking in brane-world models has several very attractive
features. The possibility of breaking supersymmetry on a distant
brane offers a geometric realization of the idea of hidden sectors.
Supersymmetry can be completely broken in a non-local way 
by partially breaking supersymmetry on different branes in such
a way that each brane preserves a different fraction of the extended 
bulk supersymmetry, e.g., \cite{antibr,Hor,Bro,intbr}. It has been 
argued that such a framework could help to tackle the cosmological 
constant problem \cite{BMQ}. In 5D orbifold models which realize this 
supersymmetry breaking scenario, 
quantitative results can be obtained for the radiatively generated mass
splittings \cite{BHN}. The exciting result is that the one-loop
contribution to the mass squared of a scalar that has Yukawa couplings
to quarks and leptons and vanishing tree-level mass is finite and 
negative.\footnote{See however \cite{GNN}, where it was shown that 
quadratic divergences arise if the gauge group is Abelian.}
A similar calculation has been performed for a non-supersymmetric
type I string model \cite{ABQ}.

In this article, we would like to study a brane-world model where
supersymmetry is not broken by some orbifold projection but rather by
the branes themselves. This is the way supersymmetry is broken in almost
all of the realistic D-brane models of string theory. A D-brane breaks 
half of the bulk supersymmetry by its mere presence. The fields of the
effective world-volume theory only fill multiplets of the smaller
supersymmetry algebra. Supersymmetry can be completely broken either 
by adding anti-D-branes, which break the half of supersymmetry that is
preserved by the D-branes, or by considering configurations of intersecting
D-branes where different intersections preserve different fractions of
the bulk supersymmetry. Such a scenario has been
called pseudo-supersymmetry \cite{pseudo}.

An effective four-dimensional description of a broad class of models where
supersymmetry is partially broken by branes was recently discussed in
\cite{BFKQ} (see also \cite{AB}). By considering only the minimal field
content, consisting of the $\cN=2$ gravity multiplet, the Goldstone fields
and their superpartners, and requiring the broken supersymmetry to be
non-linearly realized, the effective action could be uniquely determined.
Although the supergravity approach is the appropriate framework to analyze
supersymmetry breaking in brane worlds, it turns out that many important
results can already be obtained by just considering the situation in
global supersymmetry (see, e.g., \cite{MP}). 
Building upon the pioneering work of \cite{BG}, the explicit 4D effective
Lagrangian of global pseudo-supersymmetry was determined in \cite{pseudo}.
This is an important result because it applies to the low-energy
effective theory of any string model realizing the pseudo-supersymmetry 
scenario.

The aim of this work is to obtain quantitative results for the radiatively
generated mass-splittings in pseudo-supersymmetry. We consider a toy model
of two 3-branes located at $x^5=0$ and $x^5=\pi R$ in $M^{3,1}\times S^1$.
There is an $\cN_4=2$ vector corresponding to the gauge symmetry $G$ in
the bulk and $\cN_4=1$ chiral multiplets charged under the gauge symmetry
are confined to the 3-branes. The chiral multiplets from the two 3-branes 
couple to different bulk gauginos and thus preserve different halves of 
the bulk supersymmetry. Brane scalar masses are generated through Feynman 
diagrams involving a loop of fields from the distant brane. Such diagrams 
arise at the two-loop level. The explicit computation shows that the expected 
quadratic cutoff dependence is regulated by the finite brane separation. Only 
logarithmic divergences arise. They are due to wave-function renormalization
of the brane fields. More precisely, we find that the brane scalar mass 
squared is positive and, for $R$ much larger than the inverse cutoff scale
$\ms^{-1}$, proportional to $(2\pi R)^{-2}\ln(2\pi R\,\ms)$.
Similarly, we find that the Casimir energy depends quartically on the
inverse of the brane separation but only logarithmically on the cutoff scale.
Thus supersymmetry breaking is soft in this class of models.

These results are particularly interesting in view of the recent D-brane
constructions which represent embeddings of the standard model in string 
theory \cite{antibr,intbr}. The knowledge of the precise expression for
the mass splittings and their dependence on the interbrane distance is
is an important first step towards a phenomenological analysis of those
models. Although many of the explicit models have additional features,
the toy model of this article captures their main supersymmetry breaking
mechanism. The quantitative results of this article are directly applicable 
to D-brane models if the string scale is much larger than the 
compactification scale since in this limit all excited string states as
well as the states corresponding to strings stretching between the distant
D-branes are much heavier than the Kaluza-Klein excitations and can
therefore be neglected.

The paper is organized as follows. In the next section, we generalize
the results of \cite{pseudo} to five-dimensional models. Using this result,
we compute one-loop corrections to the bulk Lagrangian, two-loop corrections
to the brane Lagrangians and three-loop corrections to the vacuum energy
density. In four appendices, we explain our notation, comment on the
dimensional reduction of the 5D supersymmetry algebra, give the explicit
component Lagrangians of our model and list the Feynman rules and some
useful formulae to do the loop integrations.

\section{Pseudo-Supersymmetry in D=5}
Consider two 3-branes\footnote{In string theory, these are two stacks of
branes. Whenever writing `brane' in this article, we understand that it
could either be a single brane or a stack of branes.}
that break different halves of the bulk supersymmetry.
For concreteness, we concentrate on a flat 5-dimensional bulk space of the
form $M^{3,1}\times S^1$. The 3-branes are extended along the 4-dimensional
Minkowski space-time $M^{3,1}$ and are separated by a distance $l$ on the 
circle $S^1$ of radius $R$. For simplicity, we assume $l=\pi R$ throughout
this article. The simplest model of pseudo-supersymmetry 
consists of an $\cN_5=1$, $D=5$ vector multiplet in the bulk and $\cN_4=1$, 
$D=4$ chiral multiplets on the 3-branes. The chiral matter on each of the 
two branes couples to the bulk vector and to one of the bulk gauginos via 
the usual supersymmetric gauge couplings. The crucial point, however, is 
that the fields on the first brane couple only to the first bulk gaugino 
while the fields on the second brane couple only to the second bulk gaugino. 
The Lagrangian for the dimensional reduction of this model to $D=4$ has been 
given in \cite{pseudo}. Here, we are interested in the situation where the
radius $R$ of the fifth dimension is much larger than the inverse supersymmetry
breaking scale\footnote{This is the scale where the bulk supersymmetry is 
partially broken on the branes. In string models, $\ms$ is related to the
the string scale by $\ms=(2\pi^2 g_{\rm str})^{-1/4}\Mstr$, as has been
shown in \cite{ABL}. See also the discussion at the end of section 
\ref{two_loop_section}.} 
$\ms^{-1}$. The 5-dimensional generalization of \cite{pseudo} is easily found
using the formalism of \cite{AHGW}. To fix our notation, we review the 
Lagrangian for the 5D bulk vector multiplet, its reduction to $D=4$ and 
its form in terms of 4D superfields. Then we discuss the couplings of the
brane fields to the bulk vector.

The bulk vector multiplet contains the component fields
\be  \label{fiveD_vector}
A_M,\quad \lambda^{(5)i},\quad \phi^{(5)},\quad X^a,\qquad\quad
M=0,\ldots,3,5,\ i=1,2,\ a=1,2,3,
\ee
where $A_M$ is a 5D vector, $\lambda^{(5)i}$ is a 5D symplectic Majorana
spinor, $\phi^{(5)}$ is a real scalar and $X^a$ is an $SU(2)_R$ triplet
of real auxiliary fields. Our conventions for spinors and $\gamma$-matrices
are explained in appendix \ref{app_conv}.

The Lagrangian for this multiplet is
\bea  \label{fiveD_Lagr}
\cL_5& = &{2\over g_{(5)}^2}\tr\bigg(-\qt F^{MN}F_{MN}-\hf \DDM\phi^{(5)}
                 \DM\phi^{(5)}-{i\over2}\wbar\lambda_i^{(5)}\gamma^M \DM
                 \lambda^{(5)i}    \nonumber\\
     &&\qquad\quad +\hf X^aX^a -\hf\wbar\lambda_i^{(5)}\left[\phi^{(5)},
                    \lambda^{(5)i}\right]\bigg),
\eea
where $\cD_M=\partial_M+i[A_M,\cdot]$ denotes the covariant derivative.

The components of the $\cN_5=1$, $D=5$ vector multiplet can be rearranged
to fit into an $\cN_4=2$, $D=4$ vector multiplet
\be  \label{fourD_vector}
A_m,\quad \lambda_i,\quad \phi,\quad X^a,\qquad\quad i=1,2,\quad a=1,2,3,
\ee
where $A_m$ is a 4D vector, $\lambda_i$ are two 4D Weyl spinors,
$\phi$ is a complex scalar, and $X^a$ are three real auxiliary fields.
The precise mapping is
\be  \label{lambdaphi_mapping}
\lambda^{(5)1}=\left(\ba{c}\lambda_1\\-\bar\lambda_2\ea\right),\quad
\phi={1\over\sqrt2}\left(A_5+i\,\phi^{(5)}\right).
\ee
\quad
The $\cN_4=2$ vector can be split into an $\cN_4=1$ vector multiplet 
$V=(A_m,\lambda_1,D)$ and an $\cN_4=1$ chiral multiplet 
$\Phi=(\phi,\lambda_2,F)$. To perform this splitting, we define the
$\cN_4=1$ auxiliary fields
\be  \label{DF_mapping}
D=X^3+\cD_5\phi^{(5)},\quad D'=X^3-\cD_5\phi^{(5)},\quad 
F={i\over\sqrt2}\left(X^1+i\,X^2\right).
\ee
In appendix \ref{app_susy}, we show that $V$ and $\Phi$ transform 
irreducibly under the first supersymmetry \cite{MP}. Their superspace 
expansion reads\footnote{$V$ is in Wess-Zumino gauge and $\Phi$ in the 
$y$-basis, see \cite{WB}.}
\bea  \label{vector_split}
V &= &-\theta\sigma^m\bar\theta A_m+i\theta\theta\bar\theta\bar\lambda_1
      -i\bar\theta\bar\theta\theta\lambda_1
      +\hf\,\theta\theta\bar\theta\bar\theta D, \nonumber\\
\Phi &= &\phi+\sqrt2\,\theta\lambda_2+\theta\theta F.
\eea
Alternatively, one can split the $\cN_4=2$ vector into an $\cN_4=1$ vector 
multiplet $V'=(A_m,\lambda_2,-D')$ and an $\cN_4=1$ chiral multiplet 
$\Phi'=(\phi,-\lambda_1,F^\dagger)$ transforming irreducibly under the second
supersymmetry. Their superspace expansion reads
\bea  \label{vector_split_prime}
V' &= &-\tilde\theta\sigma^m\bar{\tilde\theta} A_m
      +i\tilde\theta\tilde\theta\bar{\tilde\theta}\bar\lambda_2
      -i\bar{\tilde\theta}\bar{\tilde\theta}\tilde\theta\lambda_2
      -\hf\,\tilde\theta\tilde\theta\bar{\tilde\theta}\bar{\tilde\theta} D', 
                                                            \nonumber\\
\Phi' &= &\phi-\sqrt2\,\tilde\theta\lambda_1
          +\tilde\theta\tilde\theta F^\dagger.
\eea
In our notation, the fermionic coordinates of superspace corresponding
to the first and second supersymmetry are denoted by $\theta$, $\bar\theta$
and $\tilde\theta$, $\bar{\tilde\theta}$ respectively.

The interesting observation of \cite{AHGW,Heb} is that the complete 5D
Lagrangian (\ref{fiveD_Lagr}) can be written in terms of 4D superfields.
Indeed, one can verify that
\be  \label{supersp_Lagr}
\cL_{\rm bulk}\ =\ {1\over2\,g_{(5)}^2}\tr\left[
                \int d^2\theta\,W^\alpha W_\alpha
               +\int d^2\bar\theta\,\bar W_{\dot\alpha}\bar W^{\dot\alpha}
               +\int d^2\theta d^2\bar\theta
                       \left(e^{-2V}\nabl e^{2V}\right)^2\right]
\ee
precisely coincides with (\ref{fiveD_Lagr}). The $x^5$-covariant derivative
\cite{Heb}
\be  \label{def_nabla}
\nabl e^{2V}=\partial_5e^{2V}+i\sqrt2\left(\Phi^\dagger e^{2V}
                                    -e^{2V}\Phi\right)
\ee
reflects the gauge invariance in the compactified dimension. Some comments
on $\nabl$ are made in the appendices \ref{app_susy} and \ref{app_comp};
for more details see \cite{Heb}.
All fields in (\ref{supersp_Lagr}) may depend on the coordinate $x^5$ and 
it is understood that the action is to be integrated over all five dimensions,
$S_{\rm bulk}=\int d^5x\,\cL_{\rm bulk}$.

Now, consider an $\cN_4=1$, $D=4$ chiral multiplet $\Phi^{(1)}$ which is
charged under the bulk gauge symmetry and localized on the first 3-brane.
For simplicity, we take $\Phi^{(1)}$ to transform in the fundamental 
representation of the gauge group. The coupling of the bulk vector to
brane matter is determined by gauge symmetry and supersymmetry.
\be  \label{brane_bulk_Lagr}
\cL^0_{(1)}=\delta(x^5)\,\int d^2\theta d^2\bar\theta\,\Phi^{(1)\dagger}
              e^{2\,V}\Phi^{(1)},
\ee
where we assumed that the first 3-brane is located at $x^5=0$.
Note that this Lagrangian contains an additional interaction between
the bulk scalar and the brane scalars of the form 
$\cD_5\I(\phi)\,\phi^{(1)\dagger}\phi^{(1)}\delta(x^5)$
as pointed out by the authors of \cite{MP}.

One of the main points of \cite{pseudo} was to show how the second
supersymmetry that is broken by the above brane-bulk coupling can still
be non-linearly realized. This is possible if the model under consideration
contains an $\cN_4=1$, $D=4$ goldstino superfield $\Lambda_g${}\footnote{%
$\Lambda_g$ is a Weyl spinor under 4D Lorentz transformations but we mostly
suppress its spinor index. The lowest component of $\Lambda_g$ is the Goldstone
fermion $\lambda_g$.} that is localized on the world-volume of the brane and 
transforms under the second supersymmetry as
\be  \label{sgstino_transf}
\delta_{\xi_2}\Lambda_g=\ms^2\,\xi_2
    -\ms^{-2}\,\Vm\partial_m\Lambda_g,
\qquad{\rm where}\quad 
\Vm=i\Lambda_g\sigma^m\bar\xi_2-i\xi_2\sigma^m\bar\Lambda_g.
\ee
The Lagrangian (\ref{brane_bulk_Lagr}) is rendered invariant under the
second supersymmetry by assigning the transformation law
\be  \label{Phi_transf}
\delta_{\xi_2}\Phi^{(1)}=-\ms^{-2}\,\Vm\partial_m\Phi^{(1)}
\ee
to the brane chiral multiplets and by adding appropriate couplings
involving the goldstino to the Lagrangian. To find these goldstino
couplings, first note that there exists a density superfield
\be  \label{Ehat_def}
\hat E\ =\ 1+{1\over8\,\ms^4}\bar D^2\bar\Lambda_g^2
            +{1\over8\,\ms^4}D^2\Lambda_g^2 + \cO\left(\ms^{-8}\right)
\ee
which has the property that $\int d^2\theta d^2\bar\theta\,
\hat E\,\Phi^\dagger\Phi$ is invariant under the second supersymmetry.
In this expression, $D_\alpha$ is the supercovariant derivative 
$\partial_\alpha+i\,(\sigma^m\bar\theta)_\alpha\partial_m$.
Then, one replaces $V$ by 
\be  \label{Vhat_def}
\hat V\ =\ V + i\ms^{-2}\,\theta\sigma^m\bar\theta\,
    (\bar\Lambda_g\bar\sigma_m D\Phi+\Lambda_g\sigma_m\bar D\Phi^\dagger)
    +\cO\left(\ms^{-4}\right)
\ee
and finds \cite{pseudo} that 
\be  \label{brane_bulk_one}
\cL_{(1)}=\delta(x^5)\,\int d^2\theta d^2\bar\theta\,\hat E\,\Phi^{(1)\dagger}
              e^{2\,\hat V}\Phi^{(1)}
\ee
is invariant under both supersymmetries.

Next, consider an $\cN_4=1$, $D=4$ chiral multiplet $\Phi^{(2)}$ localized 
on the second 3-brane and transforming irreducibly under the second 
supersymmetry. Again, the coupling of the bulk vector to brane matter is
determined by gauge symmetry and supersymmetry. But we need to split the
bulk vector vector multiplet according to (\ref{vector_split_prime}).
The Lagrangian for this brane-bulk coupling is given by
\be  \label{bulk_brane_Lagr}
\cL^0_{(2)}=\delta(x^5-l)\,\int d^2\tilde\theta d^2\bar{\tilde\theta}\,
                                   \Phi^{(2)\dagger}e^{2\,V'}\Phi^{(2)}.
\ee
This interaction is manifestly invariant under the second supersymmetry.
It can be rendered invariant under both supersymmetries by adding appropriate
goldstino interactions. Of course, there need to be a second goldstino
superfield $\Lambda_g'$ which is localized on the second 3-brane. In complete
analogy to the above definitions (\ref{Ehat_def}) and (\ref{Vhat_def}), one
defines $\hat E'$ and $\hat V'$ from $\Lambda_g'$ and $V'$. The invariant
Lagrangian is given by
\be  \label{brane_bulk_two}
\cL_{(2)}=\delta(x^5-l)\,\int d^2\tilde\theta d^2\bar{\tilde\theta}\,\hat E'\,
                           \Phi^{(2)\dagger}e^{2\,\hat V'}\Phi^{(2)}.
\ee

Finally, the total Lagrangian for our pseudo-supersymmetry toy model is the
sum of the bulk Lagrangian (\ref{supersp_Lagr}) and the two bulk-brane
interactions (\ref{brane_bulk_one}) and (\ref{brane_bulk_two}):
\bea  \label{ps_Lagr}
\cL &= &\cL_{\rm bulk}\ +\ \cL_{(1)}\ +\ \cL_{(2)} \nonumber\\
    &= &{1\over2\,g_{(5)}^2}\tr\left[
                \int d^2\theta\,W^\alpha W_\alpha
               +\int d^2\bar\theta\,\bar W_{\dot\alpha}\bar W^{\dot\alpha}
               +\int d^2\theta d^2\bar\theta
                       \left(e^{-2V}\nabl e^{2V}\right)^2\right] \\
               &&+\delta(x^5)\,\int d^2\theta d^2\bar\theta\,\hat E\,
                  \Phi^{(1)\dagger}e^{2\,\hat V}\Phi^{(1)}\ 
                 +\ \delta(x^5-l)\,\int d^2\tilde\theta d^2\bar{\tilde\theta}\,
                 \hat E'\,\Phi^{(2)\dagger}e^{2\,\hat V'}\Phi^{(2)}. \nonumber
\eea

The expansion of this Lagrangian into component fields is given in appendix 
\ref{app_comp}.

\section{Loop corrections to the bulk Lagrangian}
\begin{figure}[t]
\begin{center}
\epsffile{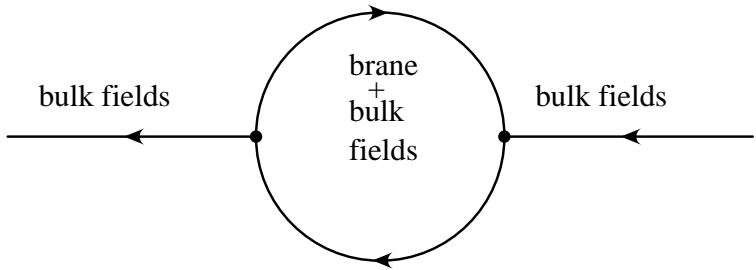}
\end{center}
\caption{\label{oneloop_diagram}One-loop Feynman diagrams contributing to
the self-energy of the bulk fields.}
\end{figure}

The $\cN_4=2$ supersymmetry of the tree-level bulk Lagrangian
(\ref{supersp_Lagr}) is broken at one-loop by the brane-bulk interactions.
Interestingly, one-loop corrections induce kinetic terms for the bulk
vector and gauginos localized on the 3-branes (for a recent discussion
of this phenomenon, see \cite{DGS}). 
The relevant diagrams contributing to the one-loop self-energy of the
bulk fields are of the form shown in fig.\ \ref{oneloop_diagram}.
Note that the goldstino couples to bulk fields only through the 
brane-bulk interactions (\ref{brane_one_comp}), (\ref{brane_two_comp}). 
As a consequence, there is no one-loop contribution to the bulk Lagrangian 
involving the goldstino.
Taking into account the residual symmetry in each subsector, one finds that
the one-loop corrected bulk Lagrangian is of the form
\bea  \label{supersp_oneloop}
\cL^{(\oneloop)}_{\rm bulk} 
               &= &{\Mc\over2\,g_0^2}\tr\left[
                \int d^2\theta\,W^\alpha W_\alpha
               +\int d^2\bar\theta\,\bar W_{\dot\alpha}\bar W^{\dot\alpha}
               +\int d^2\theta d^2\bar\theta
                     \left(e^{-2V}\nabl e^{2V}\right)^2\right] \nonumber\\
               &&+{\delta(x^5)\over2\,g_1^2}\tr\left[
                     \int d^2\theta\,W^\alpha W_\alpha
                    +\int d^2\bar\theta\,
                          \bar W_{\dot\alpha}\bar W^{\dot\alpha}\right] \\
               &&+{\delta(x^5-l)\over2\,g_2^2}\tr\left[
                     \int d^2\tilde\theta\,W^{\prime\alpha}W^\prime_\alpha
                    +\int d^2\bar{\tilde\theta}\,
                          \bar W^\prime_{\dot\alpha}\bar W^{\prime\dot\alpha}
                     \right], \nonumber
\eea
where $\Mc=(2\pi R)^{-1}$ and $W^\prime_\alpha$ is the field strength
superfield associated to the $\cN_4=1$ vector multiplet $V'$ defined in
(\ref{vector_split_prime}). The gauge couplings $g_0$, $g_1$, $g_2$ are
given by
\be  \label{gi_def}
{\Mc\over g_0^2}\ =\ {1\over g_{(5)}^2} + \Delta_{\rm bulk}, \qquad
{1\over g_i^2}\ =\ {b_i\over8\pi^2}\ln\left({m\over\Lambda}\right),
\ee
where $\Delta_{\rm bulk}$ is the contribution from the bulk fields running
in the loop, $b_1$, $b_2$ are the beta-function coefficients of the 
matter fields living on the two 3-branes, $m$ is an IR cutoff and
$\Lambda$ is a UV cutoff. We are not interested in the precise form
of $\Delta_{\rm bulk}$ but content ourselves to note that it depends
linearly on the cutoff and has been computed in \cite{DDG}. The problems
related to the non-renormalizability of the 5D Super-Yang-Mills theory
can be avoided by embedding it into a $D=5$, $\cN_5=2$ theory, i.e.,
by adding a bulk hyper multiplet in the adjoint representation. That theory
is free of ultra-violet divergences and, in particular, $\Delta_{\rm bulk}$
vanishes. 

The logarithmic divergences localized on the world-volume of the 3-branes
can be eliminated through standard four-dimensional renormalization.
Requiring the brane-localized contributions to the bulk gauge 
kinetic terms to vanish at the scale of supersymmetry breaking, $\ms$,
one has the running couplings
\be  \label{running}
{1\over g_i^2(\mu)}\ =\ {b_i\over8\pi^2}\ln\left({\mu\over\ms}\right).
\ee
It is interesting that the counterterm necessary to cancel the logarithmic
divergences is absent at tree-level and only arises through quantum 
corrections.\footnote{A model where such counterterms are already
present at tree-level, has recently been discussed in \cite{Kyae}.
In orbifold compactifications of type I string theory,
there is a tree-level contribution to brane localized gauge-kinetic terms
proportional to the expectation value of a linear combination of twisted
moduli fields \cite{IRU}.}

Supersymmetry breaking manifests itself by the different $1/g^2$ prefactors
of the various terms in the bulk Lagrangian. This is most clearly seen
in the component field expansion of (\ref{supersp_oneloop}). We find
\bea \label{bulk_oneloop}
\hf\,\cL^{(\oneloop)}_{\rm bulk} &= &
                   \left({\Mc\over g_0^2}+{\delta(x^5)\over g_1^2}
                                    +{\delta(x^5-l)\over g_2^2}\right)
                                    \tr\Bigg(-\qt F^{mn}F_{mn}\Bigg)
                                                          \nonumber\\
                 &&+\left({\Mc\over g_0^2}+{\delta(x^5)\over g_1^2}\right)
                   \tr\Bigg(
                       -{i\over2}\,\lambda_1\sigma^m\Dm\bar\lambda_1
                       -{i\over2}\,\bar\lambda_1\bar\sigma^m\Dm\lambda_1
                                     \Bigg) \nonumber\\
                 &&+\left({\Mc\over g_0^2}+{\delta(x^5-l)\over g_2^2}\right)
                   \tr\Bigg(
                       -{i\over2}\,\lambda_2\sigma^m\Dm\bar\lambda_2
                       -{i\over2}\,\bar\lambda_2\bar\sigma^m\Dm\lambda_2
                                     \Bigg) \nonumber\\
                 &&+{\Mc\over g_0^2}\tr\Bigg(-\DDm\phi^\dagger\Dm\phi
                   +F^\dagger F                              \nonumber\\
                 &&-{i\over\sqrt2}\,\eps^{ij}\left(\lambda_i[\phi^\dagger,
                       \lambda_j]+\bar\lambda_i[\phi,\bar\lambda_j]\right)
                   -\hf\eps^{ij}\left(\lambda_i\partial_5\lambda_j
                   +\bar\lambda_i\partial_5\bar\lambda_j\right)
                   +\hf (X^3)^2 \hbox{\hspace{1cm}}         \nonumber\\ 
                 &&-\hf\partial_5A^m\partial_5A_m
                   +\sqrt2\,\partial_5A^m\Dm\R(\phi)
                   -\hf\left(\sqrt2\,\partial_5\I(\phi)
                             -[\phi,\phi^\dagger]\right)^2
                   \Bigg).
\eea
The effective four-dimensional gauge coupling constant is the coefficient 
of $-\qt F^{mn}F_{mn}$ after having integrated over the fifth dimension:
\be  \label{gfour_def}
{1\over (g_{(4),\hbox{\scriptsize1-loop}})^2}\ =\ 
{1\over g_0^2}+{1\over g_1^2}+{1\over g_2^2}.
\ee
Similarly, the effective coefficients of the gaugino kinetic terms are
\be  \label{rho_def}
{1\over(\rho_{i,\hbox{\scriptsize1-loop}})^2}\ =\ 
{1\over g_0^2}+{1\over g_i^2}.
\ee
In order to have canonically normalized kinetic terms for the bulk field
zero modes after compactification to four dimensions, we have to replace
\be  \label{g_replace}
A_m\ \to\ g_{(4),\hbox{\scriptsize1-loop}}\,A_m,\quad 
\lambda_i\ \to\ \rho_{i,\hbox{\scriptsize1-loop}}\,\lambda_i,\quad
\phi\ \to\ g_0\,\phi.
\ee
This breaks supersymmetry in the brane-bulk Lagrangians
(\ref{brane_bulk_one}), (\ref{brane_bulk_two}), as can be seen from the
component Lagrangians (\ref{brane_one_comp}), (\ref{brane_two_comp}).

Although the bulk supersymmetry is broken at one-loop, there are no mass 
splittings at this order in perturbation theory. This is because each of 
the two subsectors $\cL_{\rm bulk}+\cL_{(1)}$ and $\cL_{\rm bulk}+\cL_{(2)}$
preserves $\cN_4=1$ at tree-level. Therefore the supersymmetric 
non-renormalization theorems are still valid for all diagrams involving
only fields from the bulk and one of the branes. Self-energy diagrams for
bulk fields involving couplings to fields from both branes are only possible
at three-loop. This can also be seen by considering one-loop corrections
to bulk scalar masses, which arise from diagrams of the form
$\phi\to\lambda_1\lambda_2\to\phi$. If $\lambda_1$, $\lambda_2$ were taken
to be the tree-level fields, then the bulk $\cN_4=2$ supersymmetry would
guarantee that the contribution of this diagram to the scalar mass is 
cancelled by the remaining one-loop diagrams. Only if both $\lambda_i$ 
propagators are replaced by their one-loop corrected values as in 
(\ref{g_replace}), is supersymmetry broken and no cancellation in
the mass contribution takes place. Thus, the bulk scalar acquires a mass
a three-loop.

The situation is more complicated for the fermions. Contributions to
bulk gaugino self-energy are of the form 
$\lambda_i\to\lambda_iA_m\to\lambda_i$, where the $A_m$ and $\lambda_i$ 
propagators have one-loop insertions of brane fields. This implies that 
bulk fermion masses can only arise at three-loop. However, it is easy to
see that it is not possible to obtain a non-vanishing contribution to
gaugino masses if all brane fermions are massless. In the Weyl fermion 
basis, there are two types of propagators for a fermion $\chi$ of mass $m$; 
one takes $\chi\to\bar\chi$ and is proportional to $k^m\bar\sigma_m$,
the other takes $\chi\to\chi$ and is proportional to the mass $m$.
It turns out that there is no Feynman diagram containing only Kaluza-Klein
zero modes that generates a non-vanishing gaugino mass. 
All such self-energy diagrams only represent a wave function renormalization. 
This result is the consequence of a $\IZ_4$ symmetry \cite{pseudo} 
which forbids terms quadratic in $\lambda_i$ in the Lagrangian. 
Under this symmetry, $\lambda_1$, $\lambda_2$, $\phi$, $A_m$ have charges
$1$, $-1$, $0$, $0$, respectively. However, this symmetry does not exclude
non-diagonal mass terms of the form $\lambda_1\lambda_2$. 
Indeed the excited Kaluza-Klein modes of the bulk gauginos have such
non-diagonal masses, as follows from the $\lambda_1\partial_5\lambda_2$
term in (\ref{fourD_comp}). It seems probable that Feynman diagrams
with massive Kaluza-Klein modes propagating on internal lines can generate
non-diagonal masses for the gaugino zero-modes. But an explicit calculation 
of such diagrams is beyond the scope of this article.

We would like to mention one other possibility how bulk gauginos can
acquire a mass. Gaugino masses can arise through anomaly-mediation
\cite{anom}. According to \cite{anom,Weinberg}, anomaly mediated gaugino
masses appear whenever the auxiliary fields of the gravity multiplet
acquire a vacuum expectation value (vev) and the the gauge theory has
a non-vanishing beta-function.
In section \ref{Casimir_energy}, we will see that the vacuum energy 
receives a positive contribution at three-loop. To cancel this 
vacuum energy when the model is coupled to supergravity the auxiliary
fields in the gravity multiplet acquire a vev.\footnote{We do not address
the question of how this vev arises dynamically. Here, we just assume that
there exists a mechanism to cancel the cosmological constant.} 
This vev breaks the above-mentioned $\IZ_4$ symmetry and gives mass 
to the bulk gauginos of order $(g_{(4)}^2/4\pi)^2\,\Mc^2/\Mpl$.

\section{Loop corrections to the brane Lagrangian}
\label{two_loop_section}
\begin{figure}[t]
\begin{center}
\epsffile{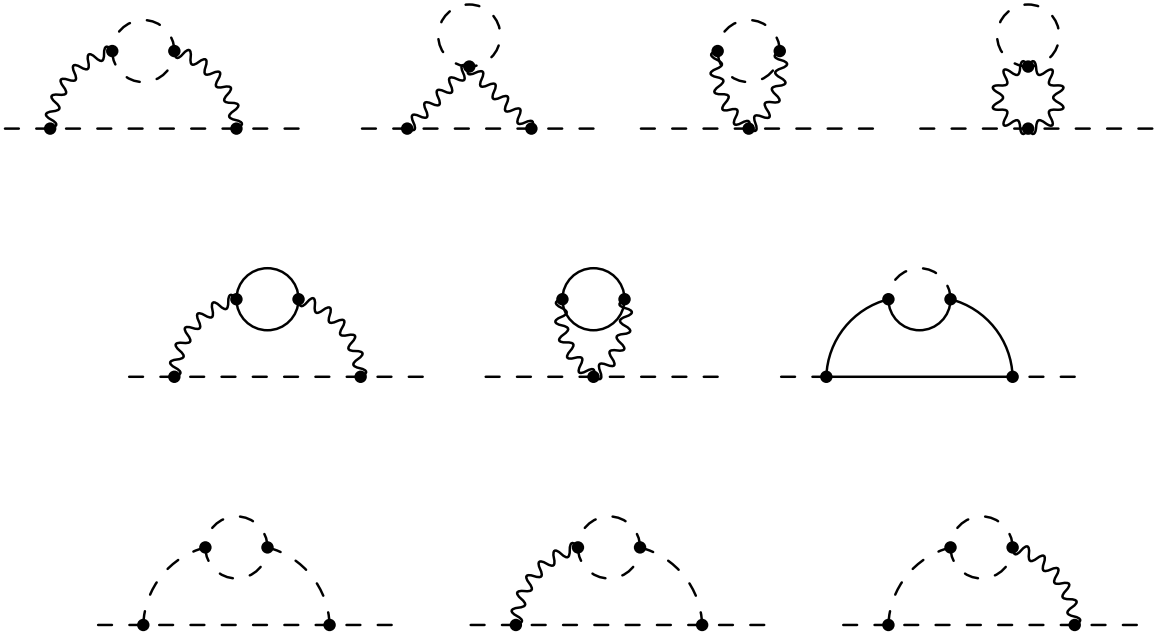}
\end{center}
\caption{\label{twoloop_diagram}Two-loop Feynman diagrams contributing to
the self-energy of the brane scalars in the case where both branes preserve
the same supersymmetry. Dashed lines are scalars, plain lines are fermions
and wavy lines are vectors. The lines in the lower level of each diagram
correspond to fields from the first brane, the lines going vertically 
correspond to bulk fields and the loop in the upper level contains fields
from the second brane. If the two branes preserve different supersymmetries,
the last diagram in the second row is absent while all other diagrams
remain unchanged.}
\end{figure}

The $\cN_4=1$ brane supersymmetry is broken at one-loop by the corrections
to the gauge coupling constant. After the replacement (\ref{g_replace}), the
$\psi^{(1)}\sigma^m\bar\psi^{(1)}A_m$ coupling is governed by 
$g_{(4),\hbox{\scriptsize1-loop}}$, eq.\ (\ref{gfour_def}), while the 
$\phi^{(1)\dagger}\lambda_1\psi^{(1)}$ coupling is governed by 
$\rho_{1,\hbox{\scriptsize1-loop}}$, eq.\ (\ref{rho_def}). However, mass
splittings only arise at two-loop. We would like to compute the two-loop
contribution to scalar masses assuming that all fields are massless to
start with. It is instructive to first consider the case where both branes
preserve the same supersymmetry, i.e., $V'$, $\hat E'$, $\tilde\theta$, 
are replaced by $V$, $\hat E$, $\theta$, respectively, in 
(\ref{ps_Lagr}).
Of course the brane scalars stay massless to all orders in perturbation
theory in this case. The relevant diagrams contributing to
the mass of the brane scalars are shown in fig.\ \ref{twoloop_diagram}. 
Supersymmetry guarantees that the sum of all these diagrams vanishes in 
the limit of vanishing external momenta. This has been explicitly verified 
in \cite{MP}.

Let us now draw the two-loop Feynman diagrams contributing to brane scalar 
masses in the case of our toy model described by the Lagrangian 
(\ref{ps_Lagr}).
Interestingly, we find exactly the same diagrams as in fig.\ 
\ref{twoloop_diagram} except for
the last diagram in the second row, which is absent because $\lambda_1$
does not couple to the fields on the second brane.\footnote{There are
several additional two-loop diagrams involving the goldstino. But they
do not contain fields from the second brane. As a consequence of the
supersymmetry preserved on the first brane, their contribution to the
scalar mass cancels.}

Let us compare in more detail the two scenarios: (i) both branes preserve 
the same supersymmetry, (ii) they preserve different supersymmetries. 
The bulk vector fields couple to brane fields in 
exactly the same way in both cases. The coupling of the bulk scalar
to the brane fields only differs by a sign in the two cases, as can
be seen from (\ref{fourD_comp}). One can verify that the last two
diagrams in fig.\ \ref{twoloop_diagram} vanish identically \cite{MP}. 
Thus, we find that the analytic expressions for all the Feynman 
diagrams in fig.\ \ref{twoloop_diagram} except for the
diagram involving $\lambda_1$ (the last diagram in the second row)
are independent of whether both branes preserve the same or a different 
supersymmetry. This observation, together with the fact that the sum of 
all diagrams cancels in the limit of vanishing external momenta in the 
first scenario, enables us to compute the sum of the nine diagrams present 
in the second scenario by just computing the single diagram diagram involving 
$\lambda_1$, which, of course, only exists in the first scenario. We conclude 
that the computation of the single diagram involving $\lambda_1$ in the case 
where both branes preserve the same supersymmetry yields the two-loop 
contribution to minus the mass squared of the brane scalars in 
pseudo-supersymmetry.

\begin{figure}[t]
\begin{center}
\epsffile{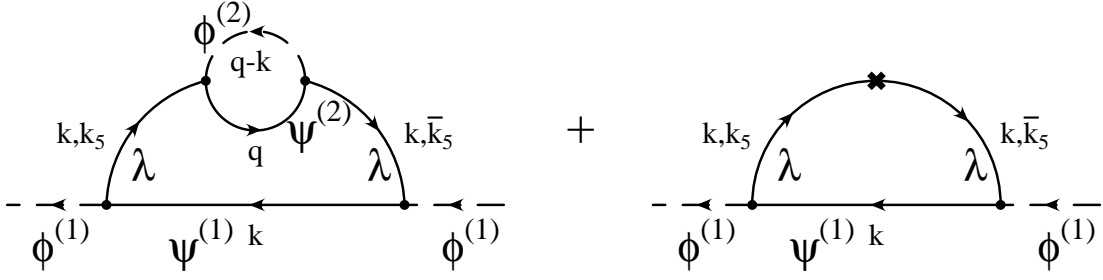}
\end{center}
\caption{\label{twoloop_detail}The leading order Feynman diagram giving
rise to brane scalar masses. This diagram does not really exist in
pseudo-supersymmetry but rather represents a shortcut to compute the
sum of the nine two-loop diagrams that do exist. The counterterm is 
fixed by the condition that the one-loop corrected bulk gauge coupling 
constant receives no contributions from brane fields at the UV cutoff 
scale $\ms$. }
\end{figure}

To do the calculation, we canonically normalize the bulk vector fields
according to the tree-level four-dimensional gauge coupling constant
$g_{(4)}=g_{(5)}(2\pi R)^{-1/2}$,
\be  \label{gtree_replace}
A_m\ \to\ g_{(4)}\,A_m,\quad 
\lambda_i\ \to\ g_{(4)}\,\lambda_i,\quad
\phi\ \to\ g_{(4)}\,\phi.
\ee
This yields the component Lagrangian (\ref{fourD_comp}). Then, we transform
the terms in this Lagrangian to the conventions where the Minkowski metric
is $\eta_{mn}=\diag(1,-1,-1,-1)$. The Feynman rules for our toy model are
shown in appendix \ref{app_formulae}. We assume that the fields on the first 
(second) brane transform according to a representation $r$ ($r'$) of the bulk 
gauge group. It turns out that the loop containing the bulk fields
in the Feynman diagram fig.\ \ref{twoloop_detail} is regularized by the
finite distance between the two branes. The loop of brane fields gives
rise to a logarithmic UV divergence which is eliminated by adding an
appropriate counterterm. 
This counterterm cannot be freely chosen but is already fixed by the
condition imposed on the bulk gauge coupling. We required that the
the contribution from the brane fields to the one-loop correction to 
the bulk gauge coupling constant should vanish at the UV cutoff scale
$\ms$. This resulted in the brane contribution to bulk coupling constant
of the form (\ref{running}) but it also fixes the precise form of the
counterterm.

\begin{figure}[t]
\begin{center}
\epsffile{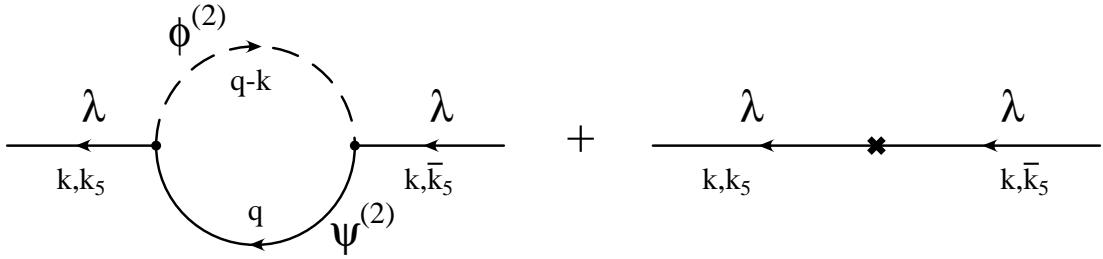}
\end{center}
\caption{\label{lambda_propag}One-loop correction to the bulk gaugino 
self-energy from fields on the second brane}
\end{figure}

To determine the precise expression of the one-loop counterterm, let us
compute the one-loop correction to the bulk gaugino self-energy from
fields on the second brane, fig.\ \ref{lambda_propag}.
The amplitude for the amputated one-loop diagram is
\be  \label{oneloop_ampl}
-i\,\Sigma_\lambda(k)\ =\ (g_{(4)}\sqrt2)^2\,d^2(r')
\int{d^4q\over(2\pi)^4}{iq^m\sigma_m\over q^2}{i\over(q-k)^2},
\ee
where $d^2(r')\,\delta^{ab}=\tr(t_{r'}^at_{r'}^b)$.
To evaluate the $q$-integral, we use dimensional regularization.
With the help of eqs.\ (\ref{Feyn_param_q}), (\ref{int_dim_reg}) and
(\ref{x_int}), we find
\be  \label{oneloop_Sigma}
-i\,\Sigma_\lambda(k)\ =\ -i\,g_{(4)}^2\,d^2(r')\,k^m\sigma_m\,
           {k^{-\eps}(4\pi)^{\eps/2}\Gamma(\eps/2)\over(4\pi)^2}
           \int_0^1dx\,[x(1-x)]^{-\eps/2},
\ee
where $\eps=4-d$. The counterterm which cancels this contribution
at an energy scale $k=\ms$ is given by
\be  \label{counterterm}
+i\,g_{(4)}^2\,d^2(r')\,k^m\sigma_m\,
           {\ms^{-\eps}(4\pi)^{\eps/2}\Gamma(\eps/2)\over(4\pi)^2}
           \int_0^1dx\,[x(1-x)]^{-\eps/2}.
\ee
For the renormalized self-energy, this yields
\be  \label{Sigma_ren}
-i\,\Sigma_{\lambda,\rm R}\ =\ \lim_{\eps\to0}\left(-i\,\Sigma_\lambda
       +\hbox{counterterm}\right)\ =\ i\,k^m\sigma_m\,
                                      {g_{(4)}^2\,d^2(r')\over8\pi^2}
                                      \ln\left({k\over\ms}\right).
\ee
This confirms our result for the running coupling, eq.\ (\ref{running}),
of the previous section. Replacing $\lambda_2\to\lambda_2/g_{(4)}$ ---
i.e., dividing $-i\,\Sigma_{\lambda,\rm R}$ by $g_{(4)}^2$ ---
and using $b_2=-d^2(r')$, (\ref{Sigma_ren}) yields the expected
correction to the kinetic terms of $\lambda_2$ as given in 
(\ref{bulk_oneloop}).

Let us now return to the computation of the mass squared of the scalars 
on the first brane arising through two-loop interactions with fields from 
the second brane. The precise expression for the Feynman diagram shown
in fig.\ \ref{twoloop_detail} is\footnote{A priori, there is an additional
contribution with $-i\,k_5\bar k_5\tr(ik^m\bar\sigma_miq^p\sigma_p)$
in the denominator. But this contribution vanishes after performing the
sum over Kaluza-Klein modes because it is odd in $k_5$, $\bar k_5$.}
\bea   \label{twoloop_expr}
-i\,m^2 &= &(-1)(-1)(g_{(4)}\sqrt2)^4\,C_2(r)d^2(r')\int_{k,q}
        {i\,\tr\left(ik^m\bar\sigma_mik^n\sigma_niq^p\bar\sigma_pik^q
                     \sigma_q\right)
         \over(q-k)^2k^2(k^2-(k_5)^2)q^2(k^2-(\bar k_5)^2)} \nonumber\\
        &&+\ (\hbox{counterterm contribution}),
\eea
where the first $(-1)$ factor is due to the fact that the diagram we are
computing is the negative of the sum of the nine two-loop diagrams that 
contribute to scalar masses and the second $(-1)$ factor is due to the
fermion loop. We have written the 5D momenta of the bulk fields in terms 
of their 4D components $(k^m, k_5)$. Note that $k_5$ need not be conserved
in a brane-bulk interaction since the branes break translational invariance
in the fifth direction.
The integration measure is given by\footnote{See \cite{MP} for a derivation
and note that the additional factors of $1/(2\pi R)$ which multiply each
of the two Kaluza-Klein sums in \cite{MP} are absent in (\ref{int_measure})
because we prefer to write all interactions in terms of the 4D gauge coupling
$g_{(4)}$. The factor $2(-1)^n$ is from the the two ways of propagating from
one brane to the other, $e^{ik_5(x^5-y^5)}+e^{ik_5(x^5+y^5)}$, where
$x^5=0$, $y^5=\pi R$ are the positions of the two branes.}
\be  \label{int_measure}
\int_{k,q}\ =\ \sum_{k_5={n\over R}}2(-1)^n
               \sum_{k_5={\bar n\over R}}2(-1)^{\bar n}
               \int{d^4k\over(2\pi)^4}\int{d^4q\over(2\pi)^4}.
\ee
To compute the sum over the Kaluza-Klein modes, we first perform a Wick
rotation and then transform the infinite sum into a contour integral in 
the complex $k_5$ plane \cite{MP,Nib}. One finds
\be  \label{KK_sum}
\sum_{k_5={n\over R}}{(-1)^n\over k^2+(k_5)^2}\,f(k)
\ =\ \oint{dk_5\over2\pi i}{\pi R\over\sin(\pi Rk_5)}
          {1\over k^2+(k_5)^2}\,f(k)
\ =\ -{\pi R\over k\sinh(\pi Rk)}\,f(k),
\ee 
where $f(k)$ is an analytic function with no poles for finite $k$.
The denominator of (\ref{twoloop_expr}) can be evaluated to give
\be  \label{denom_simpl}
i\,\tr\left(ik^m\bar\sigma_mik^n\sigma_niq^p\bar\sigma_pik^q\sigma_q\right)
    \ =\ 2i\,k^2\,k\cdot q.
\ee
To simplify the $q$-integral, we use the identity (\ref{qk_integral})
\be  \label{q_integral}
\int d^4q{k\cdot q\over q^2(q-k)^2}
\ =\ \hf k^2\,\int d^4q{1\over q^2(q-k)^2}.
\ee
Putting these results together, we find
\bea  \label{twoloop_mass}
-i\,m^2 &= &{4i\,g_{(4)}^4\over8\pi^2}\,C_2(r)d^2(r')
            (2\pi R)^2\int_0^\infty dk
            {i\,k^3\over\sinh(\pi Rk)^2}
            \int {d^4q\over(2\pi)^4}
            {1\over q^2(q-k)^2}\nonumber\\
        &&+\ (\hbox{counterterm contrib.}),
\eea  
where the additional factor $i$ in the $k$-integral is from the Wick
rotation. 

The $q$-integral can be computed in the dimensional regularization scheme 
using eqs.\ (\ref{Feyn_param}) and (\ref{int_dim_reg}).
Defining $\eps=4-d$ and adding the explicit expression for the counterterm,
which is easily derived from the Feynman diagram in fig.\ \ref{twoloop_detail}
and the result (\ref{counterterm}), the expression for the scalar mass 
squared reads
\bea  \label{twoloop_masss}
-i\,m^2 &= &{-i\,g_{(4)}^4\over2\pi^2\,(4\pi)^2}\,C_2(r)d^2(r')
            (2\pi R)^2\int_0^\infty dk
            {k^{d-1}-k^3\,\ms^{-\eps}\over\sinh(\pi Rk)^2}\nonumber\\
        &&\hspace{3cm}
            \int_0^1dx\,[x(1-x)]^{-\eps/2}\,(4\pi)^{\eps/2}\,\Gamma(\eps/2).
\eea
Finally, using the identity
\be  \label{sinh_identity}
\int_0^\infty dx\,{x^{d-1}\over\sinh(ax)^2}\ =\ 
{4\over(2a)^d}\,\Gamma(d)\,\zeta(d-1)
\ee
and taking the limit $\eps\to0$, we find
\bea  \label{scalar_mass}
m^2 &= &{g_{(4)}^4\over8\pi^4}\,C_2(r)d^2(r')
        {\Gamma(4)\zeta(3)\over(2\pi R)^2}
        \left(2\,\gamma-2{\zeta'(3)\over\zeta(3)}-{11\over3}
              +2\,\ln(2\pi R\ms)\right) \nonumber\\
    &= &\left({g_{(4)}^2\over4\pi}\right)^2 C_2(r)d^2(r')
        {24\,\zeta(3)\over\pi^2\,(2\pi R)^2}
        \,\Big(\ln(2\pi R\ms)-1.091\ldots\Big),
\eea
where $\gamma$ is the Euler constant and $\zeta'(x)={d\over dx}\zeta(x)$.
For large compactification radii, this yields a positive mass squared for
the brane scalars. For $R$ smaller than a critical radius 
$R_\star\approx0.474\,\ms^{-1}$, the potential for the brane scalars 
develops a tachyonic instability. This is very interesting since similar
tachyonic instabilities also appear in string models with D-branes and
anti-D-branes or intersecting D-branes at small separations.
Unfortunately, our approximation of neglecting the effects of states
with masses above $\ms$ is only consistent if many Kaluza-Klein modes 
have masses below $\ms$, i.e., $R\gg\ms^{-1}$.

It is interesting that the scalar masses depend only logarithmically on
the cutoff. The expected quadratic divergence is regulated
by the finite brane distance. A very similar effect has first been found
in an $S^1/(\IZ_2\times\IZ_2')$ orbifold model \cite{BHN}.\footnote{The 
authors of \cite{GNN} have shown that for an Abelian gauge group quadratic 
divergences arise in this type of orbifold models through radiatively 
generated Fayet-Iliopoulos terms. However, the analysis of \cite{GNN} 
does not apply to pseudo-supersymmetry since there is no orbifold 
projection involved in the supersymmetry breaking mechanism.}

In string theory, the UV cutoff $\ms$ is naturally identified with the
string scale. Indeed, it has been shown in \cite{ABL} that in string
models where supersymmetry is broken at the scale $\ms$ in the effective
world-volume theory on D3-branes, $\ms$ is related to the D3-brane
tension $\tau_3$ by $\hf\ms^4=\tau_3\equiv\Mstr^4/(4\pi^2 g_{\rm str})$,
where $g_{\rm str}$ is the string coupling.
The result (\ref{scalar_mass}) for the scalar masses, implies that the
mass splittings in pseudo-supersymmetric D-brane models are mainly 
determined by the compactification scale and depend only weakly on
the string scale. In particular, this could open the possibility 
to have models of intersecting D-branes where supersymmetry is broken 
at a high scale but mass splittings are only of the order of 
the much smaller compactification scale.\footnote{Here we assume
an asymmetric compactification with one large and five small compact
dimensions, which are only slightly larger than the string length
and whose effect is neglected.}
However, it is not clear whether the 5D gauge theory is consistent
at very high energy scales.
The scalar masses as a function of the compactification scale $\Mc$ for 
some specific choices the cutoff scale $\ms$ are shown in table 
\ref{smass_numbers} at the end of the next section.

\section{Casimir energy}
\label{Casimir_energy}
Since supersymmetry is broken, the vacuum energy receives non-vanishing
contributions from the quantum fluctuations of the various fields. The
dependence of this vacuum energy on the distance between the two branes
leads to a Casimir force. It is very interesting to compute the Casimir
energy because we would like to know (i) whether the Casimir force is 
attractive or repulsive and (ii) how the vacuum energy scales with the 
cutoff $\ms$.

The computation of the Casimir energy is very similar
to the computation of the mass squared of the brane scalars. It is easy 
to convince oneself that Feynman diagrams without external legs involving
fields from both branes are only possible at three-loop. For all diagrams
that contain only fields from one brane and/or the bulk, the supersymmetric
non-renormalization theorems still apply; as a consequence their contribution
to the vacuum energy cancels.

\begin{figure}[t]
\begin{center}
\epsffile{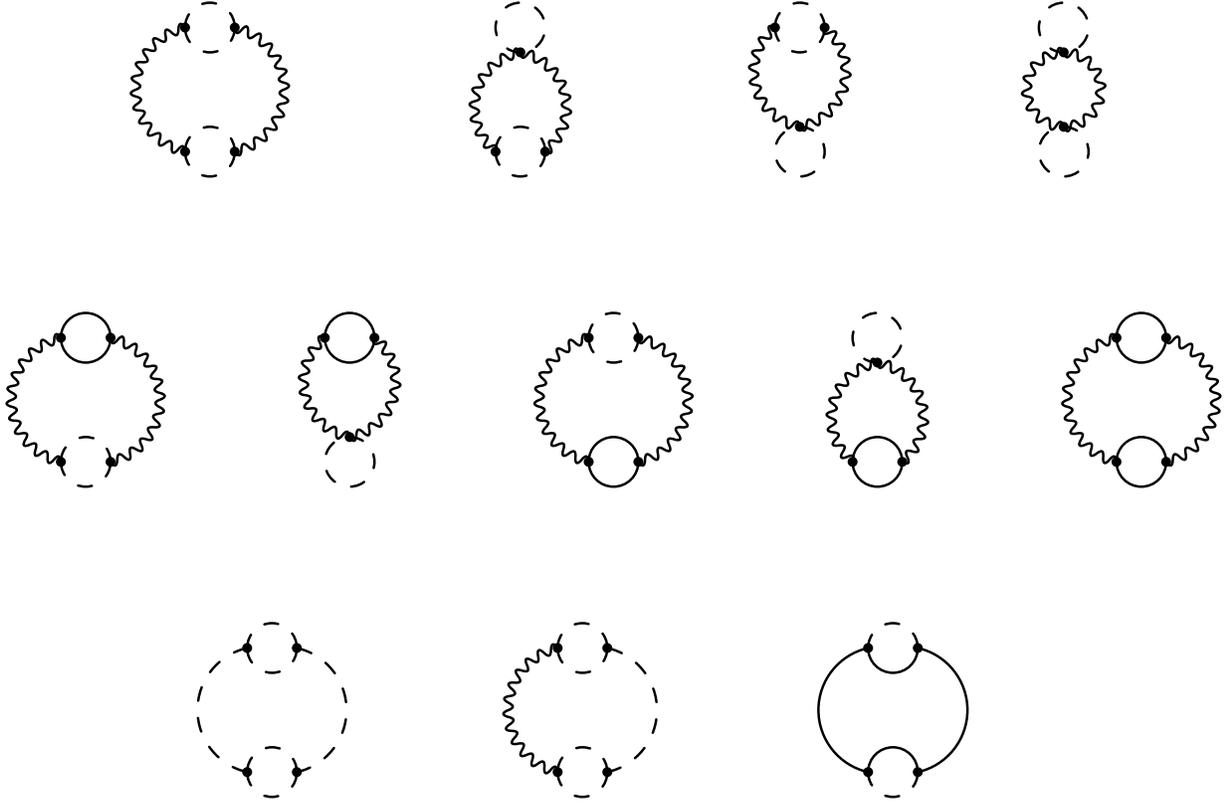}
\end{center}
\caption{\label{threeloop_diagram}Three-loop Feynman diagrams contributing to
the vacuum energy in the case where both branes preserve the same 
supersymmetry. Dashed lines are scalars, plain lines are fermions
and wavy lines are vectors. The lines in the lower level of each diagram
correspond to fields from the first brane, the lines going vertically 
correspond to bulk fields and the loop in the upper level contains fields
from the second brane. If the two branes preserve different supersymmetries,
the last diagram in the last row is absent while all other diagrams
remain unchanged.}
\end{figure}

Like in the case of the two-loop correction to the brane scalar mass, it is
useful to first consider the scenario where both branes preserve the same
supersymmetry. All three-loop diagrams without external legs involving
fields from both branes are shown in fig.\ \ref{threeloop_diagram}.

Again, one finds that only the single diagram involving $\lambda_1$ is absent
if both branes preserve different supersymmetries. A reasoning very similar
to the one of the previous section leads to the conclusion that the sum of
the eleven diagrams present in pseudo-supersymmetry can be computed by just 
computing the single diagram diagram involving $\lambda_1$, which, of course, 
only exists in the scenario where both branes preserve the same supersymmetry.

\begin{figure}[t]
\begin{center}
\epsffile{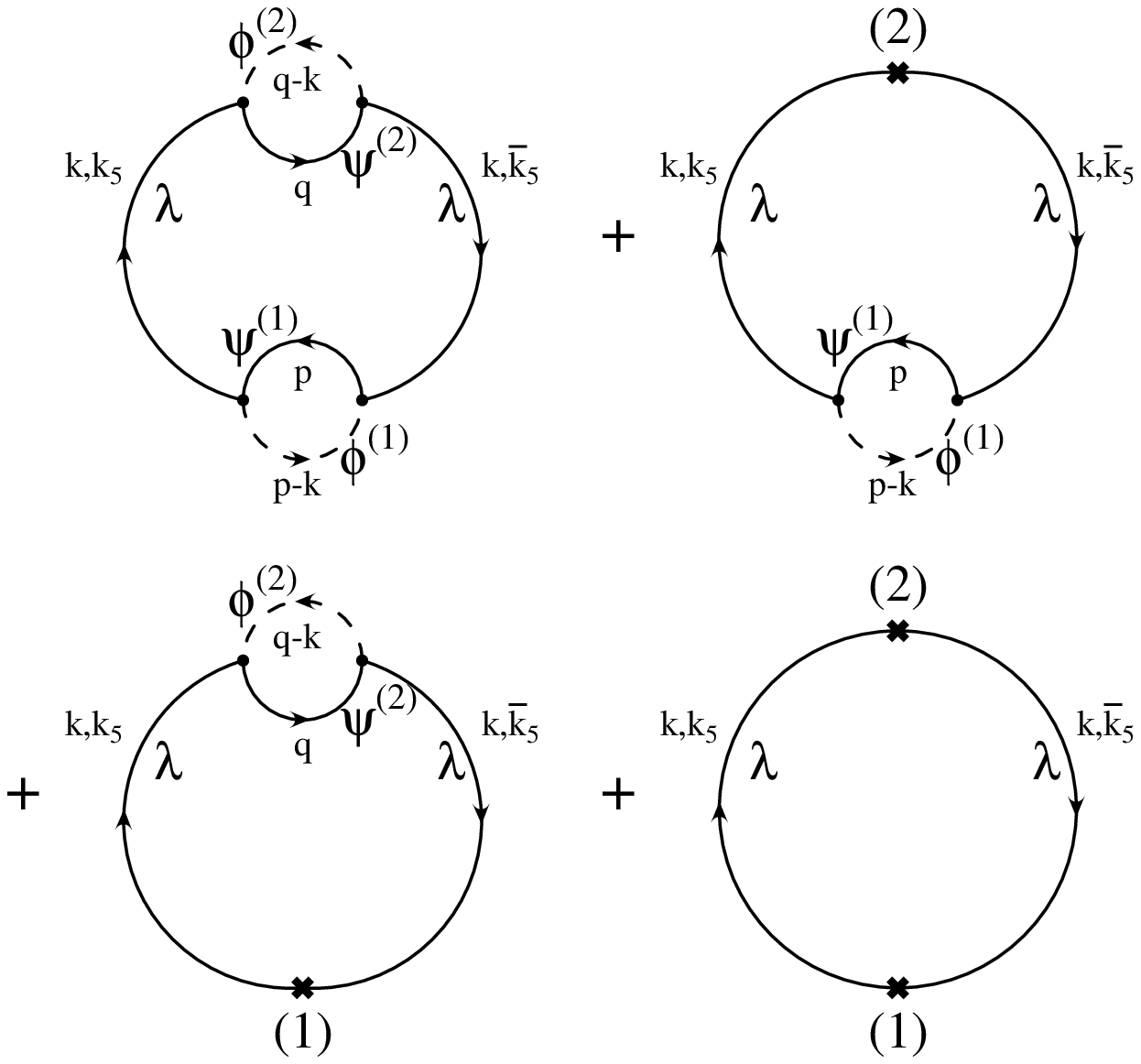}
\end{center}
\caption{\label{threeloop_detail}The leading order Feynman diagram giving
rise to a vacuum energy. This diagram does not really exist in
pseudo-supersymmetry but rather represents a shortcut to compute the
sum of the eleven three-loop diagrams that do exist. The counterterms are 
fixed by the condition that the one-loop corrected bulk gauge coupling 
constant receives no contributions from brane fields at the UV cutoff 
scale $\ms$. }
\end{figure}

In the same way as in the brane scalar mass computation, the loop-integration
over the momenta of the bulk fields in the Feynman diagram of fig.\ 
\ref{threeloop_detail} is regularized by the finite distance
between the two branes. 
The loops of brane fields give rise to logarithmic UV divergences which 
are eliminated by adding appropriate counterterms. 
Again, the counterterms are dictated by renormalization
condition imposed on the bulk gauge coupling. 

The precise expression for the Feynman diagram shown
in fig.\ \ref{threeloop_detail} is
\bea   \label{threeloop_expr}
i\,E &= &(-1)(-1)(g_{(4)}\sqrt2)^4\,d^2(r)d^2(r')\dim(G)\nonumber\\
     &&\hspace{2cm}
        \int_{k,q,p}
        {(i)^2\,\tr\left(ip^m\bar\sigma_mik^n\sigma_niq^p\bar\sigma_pik^q
                     \sigma_q\right)
         \over(q-k)^2(p-k)^2p^2(k^2-(k_5)^2)q^2(k^2-(\bar k_5)^2)} \nonumber\\
        &&+\ (\hbox{counterterm contributions}),
\eea
where the first $(-1)$ factor is due to the fact that the diagram we are
computing is the negative of the sum of the eleven three-loop diagrams that 
contribute to the vacuum energy and the second $(-1)$ factor is due to the
fermion loop. We have written the 5D momenta of the bulk fields in terms 
of their 4D components $(k^m, k_5)$. 
The integration measure is given by
\be  \label{int_measure_threeloop}
\int_{k,q,p}\ =\ \sum_{k_5={n\over R}}2(-1)^n
               \sum_{k_5={\bar n\over R}}2(-1)^{\bar n}
               \int{d^4k\over(2\pi)^4}\int{d^4q\over(2\pi)^4}
               \int{d^4p\over(2\pi)^4}.
\ee
To compute the sum over the Kaluza-Klein modes, we first perform a Wick
rotation and then transform the infinite sum into a contour integral in 
the complex $k_5$ plane \cite{MP,Nib}, as we did in the previous section,
eq.\ (\ref{KK_sum}). The denominator of (\ref{threeloop_expr}) can be
evaluated to give
\be  \label{denom_simplif} 
        (i)^2\,\tr\left(ip^m\bar\sigma_mik^n\sigma_niq^p\bar\sigma_pik^q
                     \sigma_q\right)\ =\ 
        -2\,(p\cdot k)\,(q\cdot k)+4\,(p\cdot q)\,k^2. 
\ee
The explicit expressions for the counterterms are easily derived from
their Feynman diagrams in fig.\ \ref{threeloop_detail} and the result
(\ref{counterterm}). Putting all this together and using the formulae
of appendix \ref{app_formulae}, we find
\bea  \label{threeloop_energy}
i\,E &=  &{i\,g_{(4)}^4\over4\pi^2\,(4\pi)^4}\,d^2(r)d^2(r')\dim(G)
            (2\pi R)^2\int_0^\infty dk
            {k^{5-2\eps}-2\,k^{5-\eps}\ms^{-\eps}+k^5\,\ms^{-2\eps}
             \over\sinh(\pi Rk)^2}\nonumber\\
        &&\hspace{5cm}
            \left[\int_0^1dx\,[x(1-x)]^{-\eps/2}\,(4\pi)^{\eps/2}\,
                  \Gamma(\eps/2)\right]^2.
\eea
Finally, using the identity (\ref{sinh_identity})
and taking the limit $\eps\to0$, we find
the vacuum energy density
\be  \label{vacuum_energy}
E\ =\ {4\,g_{(4)}^4\over\pi^2\,(4\pi)^4}\,d^2(r)d^2(r')\dim(G)
       {\Gamma(6)\zeta(5)\over(2\pi R)^4}
       \left(\Big(\ln(2\pi R\ms)+A\Big)^2+B\right), 
\ee
where
\bea  \label{ABC_def}
A &= &\gamma-{\zeta'(5)\over\zeta(5)}-{274\over120}
      \ \approx\ -1.679\nonumber\\
B &= &{\pi^2\over6}+{15\over4}-\left({137\over60}\right)^2
      -\left({\zeta'(5)\over\zeta(5)}\right)^2
      +{\zeta''(5)\over\zeta(5)}
      \ \approx\ 0.203
\eea
For large compactification radii, where our calculation is valid, this yields
a repulsive Casimir force. This result could have been expected from the
fact that the brane scalar masses decrease with increasing radius; as a
consequence the brane contribution to the vacuum energy decreases with
increasing radius. For infinite brane separation, one recovers the 
supersymmetric configuration with vanishing vacuum energy and mass splittings.

The non-vanishing vacuum energy for finite brane separation has two 
important implications.
First, when coupled to supergravity, it needs to be cancelled by
a non-vanishing expectation value of the auxiliary fields in the
gravity multiplet (see, e.g., \cite{Weinberg}). 
Dynamically, this vev may arise from
appropriate additional terms in the Lagrangian; but here we content
ourselves to assume that such a vev does appear. We expect that it 
implies gravitino masses of the form
\be  \label{gravitino_mass}
m_{3/2}\ =\ {\sqrt E\over\sqrt3\,\Mpl} \ =\  
   {g_{(4)}^2\over4\pi}{\Mc^2\over\Mpl}
   {\sqrt{d^2(r)d^2(r')\dim(G)\,10\,\zeta(5)}\over\pi^2}
   \sqrt{\left(\ln\left({\ms\over\Mc}\right)+A\right)^2+B}
\ee
The first equality looks very similar to the standard formula for the 
gravitino mass in spontaneously broken $\cN_4=1$, $D=4$ supergravity. 
The two differences
are: (i) the supersymmetry breaking scale is set by the radiatively 
generated vacuum energy density, not by some non-vanishing F-term, (ii)
there are two bulk gravitinos. We expect that the general formula
(\ref{gravitino_mass}) is still valid, yielding the same non-vanishing
mass for both gravitinos. The result (\ref{gravitino_mass}) is similar
to the situation in $D=4$ gauge-mediated scenarios \cite{GR} in that 
the gravitino masses are suppressed with respect to the scalar masses. 
This is not surprising since supersymmetry breaking is transmitted via
the bulk gauge fields. However, in pseudo-supersymmetry, the scalar soft
masses depend only on the compactification scale whereas in 
gauge-mediated scenarios, the scalar masses depend on ratio
of the supersymmetry breaking vev $F$ to the messenger mass scale $M$. 

Second, if the branes are D-branes, then the Casimir energy may be important
to obtain a stable configuration. In non-supersymmetric D-brane models,
the RR-force is generically attractive between the branes that preserve
different supersymmetries. The repulsive Casimir force could possibly
balance the RR-force. To understand this effect in a little more detail,
consider two D3-branes of opposite RR-charges $\mu_3$ and $-\mu_3$. 
Assume that five of the six internal dimensions are compactified at 
the string scale $\Mstr$ and one dimension is compactified at $\Mc$. 
At energies well below $\Mstr$ but above $\Mc$ the D3-branes are of 
codimension 1. The RR-force between them is constant and attractive. 
The force per unit brane-volume is given by 
$\kappa_{10}^2\mu_3^2\Mstr^5=\pi\Mstr^5$, since
$\mu_3=\sqrt\pi\kappa_{10}$. Thus, the configuration is stabilized
at
\be \label{brane_stab}
{\Mstr\over\Mc}\ \approx\
\left(
   \left({g_{(4)}^2\over4\pi}\right)^2
   d^2(r)d^2(r')\dim(G)\,{120\,\zeta(5)\over\pi^4}\,
   \ln\left({\Mstr\over\Mc}\right)^2
\right)^{1/5}.
\ee
A large ratio $\Mstr/\Mc$ is only obtained for very large numbers of 
brane fields and/or bulk gauge fields. Thus, in our simple toy model,
the Casimir force seems to be too weak to stabilize the system at
large radii where our approximation is valid. Of course, a more careful
treatment needs to take into account gravity and, in particular, address
the question how the cosmological constant is cancelled. Here, we only
wanted to show that the Casimir effect might give a relevant contribution.

It is interesting that the vacuum energy depends only logarithmically on the
cutoff. But it depends quartically on the compactification scale which has
to be chosen close to the weak scale to avoid a hierarchy problem
for scalar masses. Thus, there seems to be no natural mechanism to suppress
the cosmological constant in this model.

In table \ref{smass_numbers}, we give the explicit values of scalar masses
and Casimir energy density for some specific choices of the cutoff scale $\ms$.
\begin{table}[t]\begin{center}
\begin{tabular}{|c|cccccc|}
\hline
${\ms\over\Mc}$  &$10^2$ &$10^5$ &$10^8$ &$10^{11}$ &$10^{14}$ &$10^{17}$\\ 
        \hline\hline
$m_{\rm scalar}$ &3.205 &5.519 &7.117 &8.417 &9.541 &10.55\\ \hline   
$E^{1/4}$        &1.294 &2.359 &3.076 &3.656 &4.156 &4.601\\ \hline
\end{tabular}
\caption{\label{smass_numbers}Numeric values for brane scalar masses 
in units of $g_{(4)}^2/(4\pi)\,\sqrt{C_2(r)d^2(r')}\,\Mc$ and the 
Casimir energy density $E^{1/4}$ in units of $g_{(4)}/\sqrt{4\pi}\,
(d^2(r)d^2(r')\dim(G))^{1/4}\,\Mc$.}
\end{center}\end{table}

\section{Conclusions and outlook}
We have computed the scalar soft masses and the Casimir energy in a 5D
brane-world model where two branes preserve different halves of the bulk
supersymmetry. In this scenario, supersymmetry breaking involves no 
orbifold projection. The fields on the world-volume of the branes simply
lack the superpartners corresponding to the extended bulk supersymmetry.
In our toy model, a 5D vector multiplet propagates in the bulk and 4D
chiral multiplets charged under the bulk gauge symmetry are confined to
the branes. Supersymmetry is broken by the fact that the chiral multiplets
from the two branes couple to different bulk gauginos.

At one-loop, the gauge kinetic terms acquire contributions which are
localized on the branes. These contributions show logarithmic divergences
in the ultraviolet, which can be eliminated through standard 4D 
renormalization. 
Brane scalar masses arise at the two-loop level. It is instructive to see
in detail how radiative corrections generate a positive mass-squared for 
the brane scalars while all tree-level masses are vanishing. Interestingly,
the masses are finite, once the brane contribution to the bulk gauge coupling
constant has been renormalized by adding the appropriate counterterm.
With our choice of the renormalization condition and for large brane
separations $\pi R$, we find
$m^2\propto (2\pi R)^{-2}(\ln(2\pi R\,\ms)-1.1)$, 
where $\ms$ is an ultraviolet cutoff.
Luckily, the three-loop diagrams that contribute to the vacuum energy density
can also be calculated without too much effort. The result,
$E\propto (2\pi R)^{-4}\Big((\ln(2\pi R\,\ms)-1.7)^2+0.2\Big)$,
shows that the Casimir force is repulsive at large distances.

The main motivation for this work was the fact that there are D-brane
constructions of the standard model \cite{intbr} where supersymmetry 
is broken in the way discussed in this article. It would be very 
interesting to apply the techniques and results of this work to an
explicit intersecting brane construction of the standard model in
order to derive precise predictions for the soft masses. However, our
methods are only valid if the string scale is at least a few orders
of magnitude larger than the compactification scale. In that case,
massive string excitations can be neglected.

The computation of the soft masses enables us to compare supersymmetry
breaking in intersecting brane models (with a high enough string scale
for our approximations to be valid) to standard 4D gauge-mediation models.
In the former, the mass splittings only depend on the distance $\pi R$ 
between the intersections where the chiral multiplets are localized,
up to logarithmic corrections involving the string scale. The product
of the compactification scale $\Mc=(2\pi R)^{-1}$ and the gauge coupling
constant $g$ is the order parameter of supersymmetry breaking. In the limit 
$g\,\Mc\to0$, supersymmetry is restored; for $\Mc>\ms$, string effects become 
relevant and the field theory approximation breaks down. 
Models of gauge-mediation, in contrast, have two mass parameters: 
the vev of an auxiliary field $F$, which is the order parameter of 
supersymmetry breaking, and the mass scale $M$ of the messenger sector. 
Scalar masses squared are of order $g^4\,F^2/M^2$,
as compared to $g^4\,\Mc^2$ in pseudo-supersymmetry. Gravitino masses
are of comparable order in both scenarios, $F/\Mpl$ in gauge-mediation
and $g^2\,\Mc^2/\Mpl$ in pseudo-supersymmetry.
In gauge-mediated models, there is a clear asymmetry between the hidden
sector, where supersymmetry is broken, and the visible sector that learns
about this breaking only from interactions with the messenger sector.
In pseudo-supersymmetry, hidden and visible sector appear on equal footing. 
Each sector sees its own supersymmetry and learns from interactions with bulk
fields that the other sector does not preserve the same supersymmetry.
Actually, realistic intersecting brane constructions do not have a hidden 
sector in the usual sense. The supersymmetry breaking mechanism of these
models only requires the known fields of the standard model and their
superpartners. 

The authors of \cite{AHGLR} have developed a superfield formalism to 
determine supersymmetry breaking effects. In this formalism, soft masses 
arise through spurion superfields whose auxiliary components acquire 
vacuum expectation values $F\ll M^2$. Although this method is not 
applicable to pseudo-supersymmetry since the dimensionless small 
number $F/M^2$ has no obvious counterpart in the scenario discussed 
in this article, it is very interesting to see whether there is
some effective 4D superfield formalism that captures the radiative 
generation of soft masses in pseudo-supersymmetry.

Another direction for future work is the analysis of brane constructions
containing branes and antibranes. In those cases, only the gravity multiplet
and the moduli fields propagate in the bulk. A simple toy model consists
of an $\cN=2$ hyper multiplet in the bulk coupling with gravitational
strength to $\cN=1$ chiral multiplets on the branes. The computation of
the soft masses in that case is probably more involved because gravity
effects have to be included. However, we can get a rough estimate of
the soft terms by replacing the gauge coupling $g_{(4)}$
with the gravitational coupling $\Mc/\Mpl$ in the expression for the
brane scalar mass squared, eq.\ (\ref{scalar_mass}). Thus, we expect
brane scalar masses of order $m^2\propto\Mc^6/\Mpl^4$. Requiring the
soft masses to be of the order of 1 TeV, implies $\Mc\sim10^{13}$ GeV.
This favors a high string scale around the GUT scale.

\vskip20mm
\centerline{\bf Acknowledgements}

It is a pleasure to thank Michael Peskin for many stimulating discussions
and for proofreading the manuscript.  
I have also benefited from discussions with Adi Armoni, Thomas Becher, 
Cliff Burgess, Arthur Hebecker, Luis Ib\'a\~nez, Shamit Kachru, 
Fernando Quevedo, Sang-Jin Sin and Angel Uranga.
Special thanks go to my wife for her support and encouragement.
This research is funded by the Deutsche Forschungsgemeinschaft.

\vskip1cm

\begin{appendix}

\section{Notation and conventions}
\label{app_conv}
We use the metric $\eta_{MN}=\diag(-1,1,1,1,1)$ and a set of $\gamma$-matrices
that satisfies
\be  \label{cliff_alg}
\{\gamma^M,\gamma^N\}\ =\ -2\,\eta^{MN}.
\ee
Specifically, we take
\be  \label{gamma_matrices}
\gamma^m\ =\ \left(\ba{cc}0 &\sigma^m\\ \bar\sigma^m &0\ea\right)
\quad {\rm for}\ m=0,\ldots3\quad {\rm and}\quad
\gamma^5\ =\ \left(\ba{cc}-i\one_2 &0\\ 0 &i\one_2\ea\right),
\ee
where $\sigma^m=(-\one_2,\sigma^1,\sigma^2,\sigma^3)$,
$\bar\sigma^m=(-\one_2,-\sigma^1,-\sigma^2,-\sigma^3)$ and
\be  \label{sigma_def}
\sigma^1\ =\ \left(\ba{cc} 0 &1 \\ 1 &0\ea\right),\quad
\sigma^2\ =\ \left(\ba{cc} 0 &-i \\ i &0\ea\right),\quad
\sigma^3\ =\ \left(\ba{cc} 1 &0 \\ 0 &-1\ea\right).
\ee

A 5D symplectic Majorana spinor $\Psi^i$ satisfies the reality condition
\be  \label{Majorana_cond}
\Psi^i\ =\ i\gamma^2\,\eps^{ij}\Psi^*_j.
\ee
It can be decomposed into 4D Weyl spinors $\psi_i$ as
\be  \label{Majorana_decomp}
\Psi^i\ =\ \left(\ba{c}\psi^i\\ -\eps^{ij}\,\bar\psi_j\ea\right),\qquad
\wbar\Psi_i\ =\ \left(\eps_{ij}\psi^j,\,\bar\psi_i\right),
\ee
where $\eps^{12}=-\eps_{12}=1$. For 4D Weyl spinors, we use the conventions
of Wess and Bagger \cite{WB}. From (\ref{Majorana_decomp}), one finds
\bea  \label{Maj_Weyl_conversion}
\wbar\Psi_i\Psi^i  &= &\eps^{ij}\left(\psi_i\psi_j
                               -\bar\psi_i\bar\psi_j\right),  \\
\wbar\Psi_i(\sigma^a)^i_{\:j}\Psi^j &= &  
\psi^k\eps_{ki}(\sigma^{a\top})^i_{\:j}\psi^j
    +\bar\psi_i(\sigma^a)^i_{\:j}\eps^{jk}\bar\psi_k,
\nonumber\\
\wbar\Psi_i\gamma^M\DM\Psi^i  &= &\psi_i\sigma^m\Dm\bar\psi_i
                   +\bar\psi_i\bar\sigma^m\Dm\psi_i
                   -i\eps^{ij}\left(\psi_i\cD_5\psi_j
                               +\bar\psi_i\cD_5\bar\psi_j\right), \nonumber\\
\wbar\Psi_i(\sigma^a)^i_{\:j}\gamma^M\cD_M\Psi^j &= &
  -\psi_i(\sigma^{a\top})^i_{\:j}\,\sigma^m\cD_m\bar\psi^j
  +\bar\psi_i(\sigma^a)^i_{\:j}\,\bar\sigma^m\cD_m\psi^j \nonumber\\
 && -i\psi^k\eps_{ki}(\sigma^{a\top})^i_{\:j}\cD_5\psi^j
    -i\bar\psi_i(\sigma^a)^i_{\:j}\eps^{jk}\cD_5\bar\psi_k. \nonumber
\eea

\section{Supersymmetry algebra}
\label{app_susy}
The supersymmetry transformations of the $D=5$, $\cN_5=1$ vector multiplet
can be written in terms of four-dimensional fields without losing any
information if all derivatives with respect to $x^5$ are kept. 
It is interesting to see how the gauge invariance along the fifth
dimension manifests itself in the dimensionally reduced version of the algebra.
In $D=5$ notation, one has \cite{MP}
\bea \label{fiveD_transf}
\delta_\xi A_M &= &i\,\bar\xi_i^{(5)}\gamma_M\lambda^{(5)i}, \nonumber\\
\delta_\xi \phi^{(5)} &= &-i\,\bar\xi_i^{(5)}\lambda^{(5)i},          \\
\delta_\xi \lambda^{(5)i} &= &\left(\sigma^{MN}F_{MN}
                                   -\gamma^M \DM\phi^{(5)}\right)\xi^{(5)i}
                             +i\left(X^a\sigma^a\right)^i_{\:j}\xi^{(5)j},
                                                            \nonumber\\
\delta_\xi X^a &= &\bar\xi_i^{(5)}(\sigma^a)^i_{\:j}\gamma^M \DM\lambda^{(5)j}
                -i\left[\phi^{(5)},\,\bar\xi_i^{(5)}(\sigma^a)^i_{\:j}
                          \lambda^{(5)j}\right],            \nonumber
\eea

Inserting the definitions (\ref{lambdaphi_mapping}) and 
(\ref{gamma_matrices}), (\ref{Majorana_decomp}), this yields the 
well-known (see, e.g., \cite{West}) transformations of an
$\cN_4=2$, $D=4$ vector multiplet up to terms containing $\partial_5$.
\bea  \label{fourD}
\delta_\xi A_m &= &i\,\xi_i\sigma_m\bar\lambda_i
                   +i\,\bar\xi_i\bar\sigma_m\lambda_i, \nonumber\\
\delta_\xi \phi &= &\sqrt2\,\eps^{ij}\,\xi_i\lambda_j, \\
\delta_\xi \lambda^i &= &\sigma^{mn}\xi^iF_{mn}
                          -i\eps^{ij}\sqrt2\,\sigma^m\bar\xi_j\,
                                  \tilde\cD_m\phi
                          +\sqrt2\,\partial_5\I(\phi)\,\xi^i
                          -[\phi,\phi^\dagger]\,\xi^i
                          +i\left(X^a\sigma^a\right)^i_{\:j}\xi^j, \nonumber\\
\delta_\xi X^a &= &-\xi_i(\sigma^{a\top})^i_{\:j}\,\sigma^m\tilde\cD_m
                       \bar\lambda^j
                   +\bar\xi_i(\sigma^a)^i_{\:j}\,\bar\sigma^m\tilde\cD_m
                       \lambda^j                              \nonumber\\
               &&  +\sqrt2\,[\phi,\,\bar\xi^k\eps_{ki}(\sigma^a)^i_{\:j}\,
                       \bar\lambda^j]
                   -\sqrt2\,[\phi^\dagger,\,\xi_i(\sigma^{a\top})^i_{\:j}\,
                       \eps^{jk}\lambda_k].\nonumber
\eea
To write these transformations in a concise form, we have defined the
modified covariant derivatives
\be  \label{modif_deriv}
\tilde\cD_m\phi\equiv \cD_m\phi-{1\over\sqrt2}\partial_5 A_m,\quad
\tilde\cD_m\lambda_i\equiv\cD_m\lambda_i+{i\over4}\eps^{ij}\sigma_m
                                  \partial_5\bar\lambda_j.
\ee

Since we are interested in splitting the $\cN_4=2$ vector into two
$\cN_4=1$ multiplets, it is useful to rewrite the transformations
(\ref{fourD}) in terms of the auxiliary fields $D$, $D'$ and $F$ defined
in (\ref{DF_mapping}). We find
\bea  \label{fourD_transf}
\delta_\xi A_m &= &i\,\xi_i\sigma_m\bar\lambda_i
                   +i\,\bar\xi_i\bar\sigma_m\lambda_i, \nonumber\\
\delta_\xi \phi &= &\sqrt2\,\eps^{ij}\,\xi_i\lambda_j, \nonumber\\
\delta_\xi \lambda_1 &= &\sigma^{mn}\xi_1F_{mn}
                          +i\,\xi_1 D
                          -i\sqrt2\,\sigma^m\bar\xi_2\,\tilde\cD_m\phi
                          -\sqrt2\,\xi_2\,F^\dagger       \nonumber\\
\delta_\xi \lambda_2 &= &\sigma^{mn}\xi_2F_{mn}
                          -i\,\xi_2 D'
                          +i\sqrt2\,\sigma^m\bar\xi_1\,\tilde\cD_m\phi
                          +\sqrt2\,\xi_1\,F               \nonumber\\
\delta_\xi D &= &-\xi_1\sigma^m\cD_m\bar\lambda_1
                 +\bar\xi_1\bar\sigma^m\cD_m\lambda_1
                 +\xi_2\sigma^m\cD_m\bar\lambda_2
                 -\bar\xi_2\bar\sigma^m\cD_m\lambda_2  \\
             &&  +2\sqrt2\left([\phi,\,\bar\xi_2\bar\lambda_1]
                                -[\phi^\dagger,\,\xi_2\lambda_1]\right)
                 +2i\left(\xi_2\partial_5\lambda_1
                         -\bar\xi_2\partial_5\bar\lambda_1\right),
                                                            \nonumber\\
\delta_\xi D' &= &-\xi_1\sigma^m\cD_m\bar\lambda_1
                 +\bar\xi_1\bar\sigma^m\cD_m\lambda_1
                 +\xi_2\sigma^m\cD_m\bar\lambda_2
                 -\bar\xi_2\bar\sigma^m\cD_m\lambda_2  \nonumber\\
             &&  +2\sqrt2\left([\phi,\,\bar\xi_1\bar\lambda_2]
                                -[\phi^\dagger,\,\xi_1\lambda_2]\right)
                 +2i\left(\xi_1\partial_5\lambda_2
                         -\bar\xi_1\partial_5\bar\lambda_2\right),
                                                            \nonumber\\
\delta_\xi F &= &i\sqrt2\left(\bar\xi_1\bar\sigma^m\tilde\cD_m\lambda_2
                              -\xi_2\sigma^m\tilde\cD_m\bar\lambda_1
                              -\sqrt2\,[\phi,\,\bar\xi_1\bar\lambda_1]
                              -\sqrt2\,[\phi^\dagger,\,\xi_2\lambda_2]\right).
                                                                    \nonumber
\eea
This shows that $(A_m,\lambda_1,D)$ and $(\phi,\lambda_2,F)$ transform
as an $\cN_4=1$ vector multiplet and an $\cN_4=1$ chiral multiplet,
respectively, under the first supersymmetry. Under the second supersymmetry,
$(A_m,\lambda_2,-D')$ and $(\phi,-\lambda_1,F^\dagger)$ transform as
an $\cN_4=1$ vector multiplet and an $\cN_4=1$ chiral multiplet, respectively.

The modified covariant derivatives appearing in the transformations of
$\lambda_i$ and $F$ can be understood from five-dimensional gauge
covariance \cite{Heb}. To see this, consider the first supersymmetry
and remember that the transformation $\delta_{\xi_1}$ is composed of a
pure supersymmetry transformation followed by a gauge transformation, 
necessary to return to the Wess-Zumino gauge. The corresponding chiral 
gauge parameter superfield (in the $y$-basis) is given by
\be  \label{gauge_param}
\Lambda\ =\ 2\,\theta\sigma^m\bar\xi_1\,A_m
           -2i\theta\theta\,\bar\xi_1\bar\lambda_1.
\ee
Under a gauge transformation, the vector superfield $V=(A_m,\lambda_1,D)$
and the chiral superfield $\Phi=(\phi,\lambda_2,F)$ transform as
\be  \label{VPhi_gaugetransf}
e^{2V}\ \longrightarrow\ e^{\Lambda^\dagger}e^{2V}e^{\Lambda},\qquad
\Phi\ \longrightarrow\ e^{-\Lambda}
       \left({-i\over\sqrt2}\partial_5+\Phi\right)e^\Lambda.
\ee
The derivative term in the $\Phi$ transformation law is necessary to 
generate the gauge transformation corresponding to $A_5$. Inserting
(\ref{gauge_param}) in (\ref{VPhi_gaugetransf}), one obtains the
modified covariant derivatives (\ref{modif_deriv}) in the $\xi_1$ terms
in (\ref{fourD_transf}). An analogous reasoning applies to the second
supersymmetry. The commutator term in $\delta_\xi\lambda_i$ (more clearly
seen in (\ref{fourD})) cannot be understood using $\cN_4=1$ arguments. 
It is necessary to close the $\cN_4=2$ algebra of gauge covariant
supersymmetry transformations.

From (\ref{VPhi_gaugetransf}), one expects that \cite{Heb}
\be  \label{nabla_five}
\nabl\ =\ \partial_5+i\sqrt2\,\Phi
\ee
is a covariant derivative in the $x^5$ direction. In this expression,
$\Phi$ is understood to act according to the representation of the field
on which $\nabl$ is applied. For the vector superfield, one finds
\be \label{nablvec}
\nabl e^{2V}\ =\ \partial_5 e^{2V}+i\sqrt2\left(\Phi^\dagger e^{2V}
                                    -e^{2V}\Phi\right),
\ee
which implies
\be \label{nablvec_transf}
\nabl e^{2V}\ \longrightarrow\ e^{\Lambda^\dagger}\left(\nabl e^{2V}\right)
                               e^{\Lambda}.
\ee

\section{Component field Lagrangians}
\label{app_comp}
To obtain the expansion in component fields of the superfield Lagrangian
(\ref{ps_Lagr}), we first note the identity:
\bea  \label{eVnableV}
\hf\tr\left[\left(e^{-2V}\nabl e^{2V}\right)^2\right] &= &
2\tr\left[\left({i\over\sqrt2}\,\partial_5-\Phi^\dagger\right)e^{2V}
                  \left({i\over\sqrt2}\,\partial_5-\Phi\right)e^{-2V} 
                  +\qt\partial_5 e^{2V} \partial_5 e^{-2V}\right]\nonumber\\
&&-\tr\left(\Phi^2+\Phi^{\dagger\,2}\right).
\eea
The terms in the second line vanish under the $\int d^4\theta$ integration, 
showing that the Lagrangian (\ref{supersp_Lagr}) coincides with the expression
given in \cite{AHGW} (apart from a different definition of the auxiliary
$D$ field).

Performing the superspace integrations, we find for the component Lagrangians:
\bea  \label{bulk_comp}
\cL_{\rm bulk} &= &{1\over g_{(5)}^2}\tr\Bigg(-\qt F^{mn}F_{mn}
                   -\tilde\cD^m\phi^\dagger\tilde\cD_m\phi
                   -{i\over2}\,\lambda_i\sigma^m\tilde\cD_m\bar\lambda_i
                   -{i\over2}\,\bar\lambda_i\bar\sigma^m\tilde\cD_m\lambda_i
                   +F^\dagger F                              \nonumber\\
                 &&-{i\over\sqrt2}\,\eps^{ij}\left(\lambda_i[\phi^\dagger,
                       \lambda_j]+\bar\lambda_i[\phi,\bar\lambda_j]\right)
                   -\hf\left(\sqrt2\,\partial_5\I(\phi)-[\phi,\phi^\dagger]
                       \right)^2   +\hf (X^3)^2  \Bigg)
\eea

\bea  \label{brane_one_comp}
\cL_{(1)} &= &\delta(x^5)\,\Bigg(-\hat\DDm\phi^{(1)\dagger}\hat\Dm\phi^{(1)}
                -{i\over2}\bar\psi^{(1)}\bar\sigma^m\hat\Dm\psi^{(1)}
                -{i\over2}\psi^{(1)}\sigma^m\hat\Dm\bar\psi^{(1)}
                +F^{(1)\dagger}F^{(1)}              \nonumber\\
            &&+i\sqrt2\left(\phi^{(1)\dagger}\hat\lambda_1\psi^{(1)}
                            -\bar\psi^{(1)}\bar{\hat\lambda_1}\phi^{(1)}\right)
              +\phi^{(1)\dagger}\hat D\phi^{(1)}\Bigg)
\eea

\bea  \label{brane_two_comp}
\cL_{(2)} &= &\delta(x^5-l)\,\Bigg(
                -\hat\cD^{\prime m}\phi^{(2)\dagger}\hat\Dm'\phi^{(2)}
                -{i\over2}\bar\psi^{(2)}\bar\sigma^m\hat\Dm'\psi^{(2)}
                -{i\over2}\psi^{(2)}\sigma^m\hat\Dm'\bar\psi^{(2)}
                +F^{(2)\dagger}F^{(2)}              \nonumber\\
            &&+i\sqrt2\left(\phi^{(2)\dagger}\hat\lambda_2\psi^{(2)}
                            -\bar\psi^{(2)}\hat{\bar\lambda_2}\phi^{(2)}\right)
              -\phi^{(2)\dagger}\hat D'\phi^{(2)}\Bigg)
\eea

where $\hat\Dm=\partial_m+i\hat A_m$, $\hat\Dm'=\partial_m+i\hat A_m'$ and
\bea  \label{hat_def}
\hat A_m &= & A_m + \ms^{-2}\left(-i\,\lambda_g\sigma_m\bar\lambda_2
                                  +i\,\bar\lambda_g\bar\sigma_m\lambda_2\right)
                          \ +\ \cO\left(\ms^{-4}\right),\\
\hat\lambda_1 &= &\lambda_1 + \ms^{-2}\left(\sqrt2\,\lambda_g\,F^\dagger
                             -i\sqrt2\,\sigma^m\bar\lambda_g\dm\phi
                             -\lambda_2\,D_g+i\,\sigma^{mn}\lambda_2\,
                                                   F_{g\,mn}\right)
                          \ +\ \cO\left(\ms^{-4}\right),\nonumber\\
\hat D &= &D + \ms^{-2}\left(\!-\lambda_g\sigma^m\dm\bar\lambda_2
                            -\dm\lambda_2\sigma^m\bar\lambda_g
                            +\lambda_2\sigma^m\dm\bar\lambda_g
                            +\dm\lambda_g\sigma^m\bar\lambda_2
                            +{i\over\sqrt2}\,D_g(F-F^\dagger)\right)
                                                        \nonumber\\
        &&\qquad+\ \cO\left(\ms^{-4}\right),    \nonumber
\eea
\bea  \label{hat_prime_def}
\hat A_m' &= & A_m + \ms^{-2}\left(-i\,\lambda_g\sigma_m\bar\lambda_1
                                  +i\,\bar\lambda_g\bar\sigma_m\lambda_1\right)
                          \ +\ \cO\left(\ms^{-4}\right),\\
\hat\lambda_2 &= &\lambda_2 + \ms^{-2}\left(-\sqrt2\,\lambda_g\,F
                             +i\sqrt2\,\sigma^m\bar\lambda_g\dm\phi
                             -\lambda_1\,D_g+i\,\sigma^{mn}\lambda_1\,
                                                   F_{g\,mn}\right)
                          \ +\ \cO\left(\ms^{-4}\right),\nonumber\\
\hat D' &= &D' + \ms^{-2}\left(\lambda_g\sigma^m\dm\bar\lambda_1
                            +\dm\lambda_1\sigma^m\bar\lambda_g
                            -\lambda_1\sigma^m\dm\bar\lambda_g
                            -\dm\lambda_g\sigma^m\bar\lambda_1
                            +{i\over\sqrt2}\,D_g(F-F^\dagger)\right)
                                                        \nonumber\\
        &&\qquad+\ \cO\left(\ms^{-4}\right).    \nonumber
\eea

To obtain these results, we assumed that the goldstino resides in an
$\cN_4=1$ Maxwell multiplet
\be  \label{Lambdag_def}
\Lambda_{g\,\alpha}\ =\ \lambda_g + \hf\,\theta_\alpha\,D_g
                 -{i\over2}\left(\sigma^{mn}\theta\right)_\alpha F_{g\,mn}
                 +i\,\theta\sigma^m\bar\theta\,\dm\lambda_{g\,\alpha}
                 +\,\ldots
\ee
Note that the correctly normalized superpartner of the $U(1)$ gauge field
strength $F_{g\,mn}$ is $\lambda_g'=2i\,\lambda_g$. The normalization in
(\ref{Lambdag_def}) is such that $\delta_{\xi_2}\Lambda_g=\ms^2\,\xi_2
+\cO(\ms^{-4})$. The factor $\hf$ in $\Lambda_{g\,\alpha}=\hf W_{g\,\alpha}$
is necessary for the non-linear action for $W_{g\,\alpha}$ to reproduce the
Born-Infeld action for the $U(1)$ gauge field \cite{BG,pseudo}.

Replacing $A_m$, $\lambda_i$, $D$ $\to$ $g_{(4)}A_m$, $g_{(4)}\lambda_i$,
$g_{(4)}D$, integrating out the auxiliary fields $X^3$, $F$, $F^\dagger$, 
$D_g$, $F^{(i)}$, $F^{(i)\dagger}$ and using the relation 
$g_{(5)}^2=2\pi R\,g_{(4)}^2$ one finds
\bea  \label{fourD_comp}
\cL &= &{1\over2\pi R}\tr\Bigg(-\qt F^{mn}F_{mn}
              -\tilde\DDm\phi^\dagger\tilde\Dm\phi
              -{i\over2}\,\lambda_i\sigma^m\tilde\Dm\bar\lambda_i
              -{i\over2}\,\bar\lambda_i\bar\sigma^m\tilde\Dm\lambda_i
                                               \nonumber\\
            &&-{ig_{(4)}\over\sqrt2}\,\eps^{ij}\left(\lambda_i[\phi^\dagger,
                       \lambda_j]+\bar\lambda_i[\phi,\bar\lambda_j]\right)
              -\hf\left(\sqrt2\,\partial_5\I(\phi)-[\phi,\phi^\dagger]
                       \right)^2
                   \Bigg)\nonumber\\[1ex]
            &&+\delta(x^5)\,\Bigg(-\hat\DDm\phi^{(1)\dagger}\hat\Dm\phi^{(1)}
                -{i\over2}\bar\psi^{(1)}\bar\sigma^m\hat\Dm\psi^{(1)}
                -{i\over2}\psi^{(1)}\sigma^m\hat\Dm\bar\psi^{(1)}   \\
            &&+i\sqrt2\,g_{(4)}\left(\phi^{(1)\dagger}\hat\lambda_1\psi^{(1)}
                            -\bar\psi^{(1)}\bar{\hat\lambda_1}\phi^{(1)}\right)
              +g_{(4)}\,\phi^{(1)\dagger} \bar D \phi^{(1)}\Bigg)\nonumber\\
            &&+\delta(x^5-l)\,\Bigg(
                -\hat\cD^{\prime m}\phi^{(2)\dagger}\hat\Dm'\phi^{(2)}
                -{i\over2}\bar\psi^{(2)}\bar\sigma^m\hat\Dm'\psi^{(2)}
                -{i\over2}\psi^{(2)}\sigma^m\hat\Dm'\bar\psi^{(2)}
                                               \nonumber\\
            &&+i\sqrt2\,g_{(4)}\left(\phi^{(2)\dagger}\hat\lambda_2\psi^{(2)}
                            -\bar\psi^{(2)}\bar{\hat\lambda_2}\phi^{(2)}\right)
              -g_{(4)}\,\phi^{(2)\dagger} \bar D' \phi^{(2)}\Bigg) \nonumber\\
            &&-\hf\,g_{(4)}^2\left(\phi^{(1)\dagger}\phi^{(1)}\right)^2
               \delta(x^5)^2
              -\hf\,g_{(4)}^2\left(\phi^{(2)\dagger}\phi^{(2)}\right)^2
               \delta(x^5-l)^2 \ +\ \cO\left(\ms^{-4}\right)\nonumber
\eea
where
\bea  \label{Dbar_def}
\bar D &= &\sqrt2\,\partial_5\I(\phi)-[\phi,\phi^\dagger] \nonumber\\
       &&  +\ms^{-2}\left(-\lambda_g\sigma^m\dm\bar\lambda_2
                          -\dm\lambda_2\sigma^m\bar\lambda_g
                          +\lambda_2\sigma^m\dm\bar\lambda_g
                          +\dm\lambda_g\sigma^m\bar\lambda_2\right)
           +\cO\left(\ms^{-4}\right)\nonumber\\
\bar D' &= &\sqrt2\,\partial_5\I(\phi)-[\phi,\phi^\dagger] \\
       &&  +\ms^{-2}\left(\lambda_g\sigma^m\dm\bar\lambda_1
                          +\dm\lambda_1\sigma^m\bar\lambda_g
                          -\lambda_1\sigma^m\dm\bar\lambda_g
                          -\dm\lambda_g\sigma^m\bar\lambda_1\right)
           +\cO\left(\ms^{-4}\right)\nonumber
\eea
and $\hat\lambda_i$ are as given in (\ref{hat_def}), (\ref{hat_prime_def})
but with $D_g$ and $F$, $F^\dagger$ set to zero.

\section{Feynman rules and loop-integrals}
\label{app_formulae}
To compute the Feynman diagrams, we convert the Lagrangian (\ref{fourD_comp})
to the conventions where the Minkowski metric is $\eta_{mn}=\diag(1,-1,-1,-1)$.

The propagator for a 4D Dirac spinor is
\be  \label{Dirac_prop}
\fpropagator{\Psi}{\bar\Psi}
\quad\equiv\ \contraction{\Psi}{\bar\Psi}
\ =\ {i\over p^2-m^2}\left(\gamma^mp_m+\one_4m\,\right).
\ee
This can be decomposed into its Weyl spinor components. Using
(\ref{gamma_matrices}) and
\be  \label{Dirac_Weyl}
\Psi=\left(\ba{c}\psi_1 \\ \bar\psi_2\ea\right),\qquad
\bar\Psi=\left(\psi_2,\,\bar\psi_1\right),
\ee
one finds
\be  \label{Weyl_contract}
\contraction{\Psi}{\bar\Psi}\ =\ 
{i\over p^2-m^2}\left(\ba{cc} m &\sigma^mp_m\\ \bar\sigma^mp_m &m\ea\right)
\ \equiv \ \left(\ba{cc}\contraction{\psi_1}{\psi_2} 
                  &\contraction{\psi_1}{\bar\psi_1}\\ 
                   \contraction{\bar\psi_2}{\psi_2} 
                  &\contraction{\bar\psi_2}{\bar\psi_1}\ea\right).
\ee 
Thus, the Weyl spinor propagators are given by
\bea  \label{Weyl_prop}
\fpropagator{\psi_1}{\bar\psi_1} \quad =\quad {i\,\sigma^mp_m\over p^2-m^2},
&\qquad\quad &
\fpropagator{\bar\psi_2}{\psi_2} \quad =\quad 
                           {i\,\bar\sigma^mp_m\over p^2-m^2}, \nonumber\\
\fpropagator{\psi_1}{\psi_2} \quad =\quad {i\,m\over p^2-m^2},
&\qquad\quad &
\fpropagator{\bar\psi_2}{\bar\psi_1} \quad =\quad {i\,m\over p^2-m^2}.
\eea
The propagator for a scalar field is
\be  \label{scalar_prop}
\fpropagator{\phi}{\phi^\dagger} \quad =\quad {i\over p^2-m^2}.
\ee
For the brane fields $\phi^{(i)}$, $\psi^{(i)}$, we use the above propagators
with $m=0$. For the bulk fields, we have to sum over all Kaluza-Klein modes
with masses $m_n={n\over R}$, $n=0,\ldots,\infty$.

The only vertex that we need for our computation is
\be  \label{philampsi}
\fvertex{\psi^{(i)}}{\lambda_1^a}{\phi^{(i)\dagger}}\quad =\quad
g_{(4)}\,t_r^a\,\sqrt2,
\ee
where we assumed that the brane fields $\phi^{(i)}$, $\psi^{(i)}$ transform
according to a representation $r$ of the gauge group $G$ and $t_r^a$,
$a=1,\ldots,\dim(G)$, are the gauge group generators in this representation.

Our conventions for the invariants that can be formed out of the
representation matrices are
\be  \label{invariants_def}
t_r^at_r^a\ =\ C_2(r)\,\one_{\dim(r)},\qquad
\tr(t_r^at_r^b)\ =\ d^2(r)\,\delta^{ab}.
\ee 
The normalization of the Lagrangians in this article is chosen such that
$d^2({\rm fund})=\hf$.

The following identities are useful to do the loop-integrations:
\be  \label{Feyn_param}
\int{d^dq\over(2\pi)^d}{1\over q^2(q-k)^2}\ =\ 
\int_0^1dx\int{d^dq\over(2\pi)^d}{1\over[q^2+x(1-x)k^2]^2},
\ee

\bea  \label{Feyn_param_q}
\int{d^dq\over(2\pi)^d}{q^m\over q^2(q-k)^2} &= & 
\int_0^1dx\int{d^dq\over(2\pi)^d}{(q+xk)^m\over[q^2+x(1-x)k^2]^2} \nonumber\\
&= &k^m\int_0^1dx\int{d^dq\over(2\pi)^d}{x\over[q^2+x(1-x)k^2]^2},
\eea

\be  \label{qk_integral}
\int{d^dq\over(2\pi)^d}{q\cdot k\over q^2(q-k)^2}
\ =\ \int{d^dq\over(2\pi)^d}{-\hf(q-k)^2+\hf q^2+\hf k^2\over q^2(q-k)^2}
\ =\ \hf k^2\,\int{d^dq\over(2\pi)^d}{1\over q^2(q-k)^2},
\ee

\be  \label{int_dim_reg}
\int{d^dq\over(2\pi)^d}{1\over[q^2+x(1-x)k^2]^2}\ =\ 
{i\over(4\pi)^{d/2}}{\Gamma(2-d/2)\over[x(1-x)k^2]^{2-d/2}},
\ee

\be  \label{x_int}
\int_0^1dx\,x[x(1-x)]^\alpha\ =\ \hf\int_0^1dx\,[x(1-x)]^\alpha,
\ee

\be  \label{logx_int}
\int_0^1dx\,\ln\Big(x(1-x)\Big)\ =\ -2.
\ee

To find the $\eps\to0$ limit of the dimensionally regularized loop-integrals,
we need the $\eps$-expansions
\bea  \label{eps_expans}
A^\eps &= &1+\eps\,\ln(A)+\hf\,\eps^2\,\ln(A)^2 
            \ +\ \cO\left(\eps^3\right), \\
\Gamma(\eps/2) &= &{2\over\eps}-\gamma
            +\eps\,\left({\pi^2\over24}+{\gamma^2\over4}\right)
            \ +\ \cO\left(\eps^2\right),\\
\Gamma(d-\eps) &= &\Gamma(d)\left(1-\eps\,\psi(d)
            +\hf\,\eps^2\,\left(\psi'(d)+\psi(d)^2\right)
            \ +\ \cO\left(\eps^3\right)\right),\\
\zeta(d-\eps) &= &\zeta(d)\left(1-\eps{\zeta'(d)\over\zeta(d)}
            +\hf\,\eps^2\,{\zeta''(d)\over\zeta(d)}
            \ +\ \cO\left(\eps^3\right)\right),
\eea
where 
\bea  \label{psi_def}
\psi(x)  &= &{\Gamma'(x)\over\Gamma(x)}
         \ =\ \sum_{n=1}^{x-1}{1\over n}-\gamma, \quad
         {\rm for\ }x\in\IN,\\
\psi'(x) &= &{d\over dx}\psi(x), \qquad\quad
\zeta'(x)\ =\ {d\over dx}\zeta(x).
\eea

\end{appendix}

\end{document}